\documentclass[journal]{vgtc}                     % final (journal style)
\onlineid{0}

\vgtccategory{Research}

\vgtcpapertype{please specify}

\title{Virtual Memory for 3D Gaussian Splatting}

\author{%
  \authororcid{Jonathan Haberl}{0009-0001-9502-4640},
  \authororcid{Philipp Fleck}{0000-0002-9827-2594}, and 
  \authororcid{Clemens Arth}{0000-0001-6949-4713}
}

\authorfooter{
  \item
  	Jonathan Haberl, Philipp Fleck, and Clemens Arth are with Graz University of Technology.
  	E-mail: jonathan.haberl@alumni.tugraz.at, \{philipp.fleck | clemens.arth\}@tugraz.at
}

\abstract{%
  3D Gaussian Splatting represents a breakthrough in the field of novel view synthesis. It establishes Gaussians as core rendering primitives for highly accurate real-world environment reconstruction. Recent advances have drastically increased the size of scenes that can be created. In this work, we present a method for rendering large and complex 3D Gaussian Splatting scenes using virtual memory. By leveraging well-established virtual memory and virtual texturing techniques, our approach efficiently identifies visible Gaussians and dynamically streams them to the GPU just in time for real-time rendering. Selecting only the necessary Gaussians for both storage and rendering results in reduced memory usage and effectively accelerates rendering, especially for highly complex scenes. Furthermore, we demonstrate how level of detail can be integrated into our proposed method to further enhance rendering speed for large-scale scenes. With an optimized implementation, we highlight key practical considerations and thoroughly evaluate the proposed technique and its impact on desktop and mobile devices.
}


\graphicspath{{figs/}{figures/}{pictures/}{images/}{./}} % where to search for the images

\usepackage{amsmath}
\usepackage{letltxmacro}
\usepackage{xspace}
\renewcommand{\vector}[1]{\begin{pmatrix} #1 \end{pmatrix}}

\newcommand{\mat}[1]{\ensuremath{\mathbf{#1}}}
 
\newcommand{\dotp}[2]{<#1,#2>}
\newcommand{\ident}{\mat{I}}                     % identity matrix
\newcommand{\laplace}{\ensuremath{\triangle}}                                            % laplace operator
\newcommand{\kroneker}{\ensuremath{\delta_\vartriangle}}                                            % laplace operator
\newcommand{\conj}[1][non]{\ensuremath{ \ifthenelse{\equal{#1}{non}}{\ast}{#1^\ast} } }                                            % conjugate complex
\newcommand{\conv}{\ensuremath{\ast}}
\newcommand{\degree}[1][non]{\ensuremath{\ifthenelse{\equal{#1}{non}}{^\circ}{#1^\circ}}}
\newcommand{\I}{\ensuremath{j}}

\newcommand{\FT}[1][non]{\ensuremath{\mathfrak F\ifthenelse{\equal{#1}{non}}{}{ \left\{ #1 \right\}}}}
\newcommand{\IFT}[1][non]{\ensuremath{\mathfrak F^{-1}\ifthenelse{\equal{#1}{non}}{}{ \left\{ #1 \right\} }} }
\def\cos{\mathop{\rm cos}}
\def\sin{\mathop{\rm sin}}

\newcommand{\atan}[1][non]{\arctan\ensuremath{\ifthenelse{\equal{#1}{non}}{}{ \left( #1 \right)}}}
\newcommand{\E}[1][non]{\ensuremath{\ifthenelse{\equal{#1}{non}}{E}{E\left\{ #1 \right\}}}}
\newcommand{\N}[3][]{\ensuremath{\mathcal{N}^{#1}(#2,#3)}} % normal distribution
\newcommand{\chis}[2][non]{\ensuremath{\chi^2 \ifthenelse{\equal{#1}{non}}{}{ \left(#1,#2\right)}}}
\newcommand{\CN}[3][]{\ensuremath{\mathcal{N}_\circ^{#1}(#2,#3)}} % von Mises distribution, circular normal
\newcommand{\R}[2][xx]{\ensuremath{R_{#1}\ifthenelse{\equal{#2}{}}{}{\left( #2 \right)}}} % correlation
\renewcommand{\S}[2][xx]{\ensuremath{S_{#1}\ifthenelse{\equal{#2}{}}{}{\left( #2 \right)}}} % spectral density
\newcommand{\Gammaf}[1][non]{\ensuremath{\Gamma\ifthenelse{\equal{#1}{non}}{}{ \left( #1 \right)}}}
\newcommand{\listofsymbols}[1][non]{
\ifthenelse{\equal{#1}{non}}{{\noindent \Large \textbf{List of Symbols}}}{#1}\\
\begin{longtable}{p{3cm} l}
$\dotp{x}{y}$ & Tensor dot product \\
$\dot{x}(t)$  & Derivative with respect to $t$\\ %\deri{x}{t} \\
\E        & Expectation value \\
\FT       & Fourier transform \\
\IFT      & Inverse Fourier transform \\
\I        & Imaginary number, $\sqrt{-1}$ \\
$\Im$     & Imaginary part of a complex number \\
$\Re$     & Real part of a complex number \\
$\delta$  & Dirac delta function \\
\kroneker & Kroneker delta function \\
x \conv\ y     & Convolution \\
\conj[x]  & Conjugate complex \\
$\nabla$  & Nabla operator \\
\laplace  & Laplace operator \\
\ident    & Identity matrix \\
\N{\mu}{\sigma}   & Normal distribution \\
\CN{\mu}{\kappa}  & Von Mises distribution (circular normal distribution) \\
\chis[n]{\sigma}  & Chi-square distribution\\
$\Gamma(\alpha)$          & Gamma function \\
\R{\tau}  & Autocorrelation \\
\S{i\omega} & Spectral density function \\
\end{longtable}}

\LetLtxMacro{\oldsqrt}{\sqrt} % makes all sqrts closed
\renewcommand{\sqrt}[1][\ ]{%
  \def\DHLindex{#1}\mathpalette\DHLhksqrt}
\def\DHLhksqrt#1#2{%
  \setbox0=\hbox{$#1\oldsqrt[\DHLindex]{#2\,}$}\dimen0=\ht0
  \advance\dimen0-0.2\ht0
  \setbox2=\hbox{\vrule height\ht0 depth -\dimen0}%
  {\box0\lower0.71pt\box2}}

\makeatletter
\DeclareRobustCommand\onedot{\futurelet\@let@token\@onedot}
\def\@onedot{\ifx\@let@token.\else.\null\fi\xspace}
\makeatother

\def\eg{\emph{e.g.}\xspace}

\def\etal{\emph{et al}\onedot}
\usepackage{balance}
\usepackage{comment}
\usepackage[]{glossaries}

\setglossarysection{paragraph}
\makeglossaries
\newacronym{vh}{VH}{Visual Hull}
\newacronym{dof}{{DOF}}{Degrees of Freedom}
\newacronym{ptp}{{PTP}}{Perspective 3 Point}
\newacronym{dog}{{DoG}}{Difference of Gaussian}
\newacronym{fov}{{FoV}}{Field Of View}
\newacronym{roi}{{ROI}}{Region Of Interest}
\newacronym{tv}{{TV}}{Total Variation}
\newacronym{dsc}{{DSC}}{Dice Similarity Coefficient}
\newacronym{hr}{{HR}}{Hit Rate}
\newacronym{far}{{FAR}}{False Alarm Rate}
\newacronym{ap}{{AP}}{Affinity Propagation}
\newacronym{tof}{{ToF}}{Time of Flight}
\newacronym{pf}{{PF}}{Particle Filter}
\newacronym{sfs}{{SfS}}{Shape from Silhouette}
\newacronym{sfm}{{SfM}}{Structure from Motion}
\newacronym{mp}{MP}{Mega Pixel}
\newacronym{ibvh}{IBVH}{Image Based Visual Hull}
\newacronym{ism}{ISM}{Implicit Shape Model}
\newacronym{icp}{ICP}{Iterative Closest Point}
\newacronym{sl}{SL}{Structured Light}
\newacronym{gac}{GAC}{Geodesic Active Contours}
\newacronym{gmm}{GMM}{Gaussian Mixture Models}
\newacronym{em}{EM}{Expectation Maximization}
\newacronym{cpu}{CPU}{Central Processing Unit}
\newacronym{gpu}{GPU}{Graphics Processing Unit}
\newacronym{sir}{SIR}{Sequential Importance Re-Sampling}
\newacronym{ct}{CT}{Computer Tomography}
\newacronym{mri}{MRI}{Magnetic Resonance Imaging}
\newacronym{gps}{GPS}{Global Positioning System}
\newacronym{bob}{BOB}{Bag of Boundaries}
\newacronym{cad}{CAD}{Computer Aided Design}
\newacronym{sota}{SotA}{State-of-the-Art}
\newacronym{3dgs}{3DGS}{3D Gaussian Splatting}
\newacronym{nvs}{NVS}{Novel View Synthesis}
\newacronym{lod}{LOD}{Level of Detail}
\newacronym{nerf}{NeRF}{Neural Radience Fields}
\newacronym{ply}{PLY}{Polygon File Format}
\newacronym{gltf}{glTF}{GL Transmission Format 2.0}
\newacronym{glsl}{GLSL}{OpenGL Shading Language}
\newacronym{cuda}{CUDA}{Compute Unified Device Architecture}
\newacronym{rle}{RLE}{Run-Length Encoding}
\newacronym{mlp}{MLP}{Multi-layer Perceptron}
\newacronym{sdf}{SDF}{Signed Distance Function}
\newacronym{ddr}{DDR3}{Double Data Rate 3.0}
\newacronym{pcie}{PCIe}{Peripheral Component Interconnect Express}
\newacronym{msvc}{MSVC}{Microsoft Visual C++}
\newacronym{fps}{FPS}{Frames per Second}
\newacronym{psnr}{PSNR}{Peak Signal-to-Noise Ratio}
\newacronym{ssim}{SSIM}{Structural Similarity Index Measure}
\newacronym{uav}{UAV}{Unmanned Aerial Vehicle}
\newacronym{soc}{SoC}{System-On-Chip}

\usepackage{listings}
\usepackage{subcaption}
\usepackage{multirow}

\usepackage{mathptmx}                  % use matching math font

\newcommand{\pf}[1]{\textcolor{blue}{#1}}

\newcommand{\precapspace}[0]{\vspace{-10pt}}
\newcommand{\postcapspace}[0]{\vspace{-15pt}}

\begin{document}

%%%%%%%%%%%%%%%%%%%%%%%%%%%%%%%%%%%%%%%%%%%%%%%%%%%%%%%%%%%%%%%%
%%%%%%%%%%%%%%%%%%%%%% START OF THE PAPER %%%%%%%%%%%%%%%%%%%%%%
%%%%%%%%%%%%%%%%%%%%%%%%%%%%%%%%%%%%%%%%%%%%%%%%%%%%%%%%%%%%%%%%

\firstsection{Introduction}

\maketitle

%% \section{Introduction} %for journal use above \firstsection{..} instead
% !TEX root = main.tex
%\chapter{Introduction}
\label{chap:introduction}

% General topic and background

\begin{figure}[t]
    \centering
    \vspace{-10pt}
    \includegraphics[width=\linewidth,trim=0 20 0 40,clip]{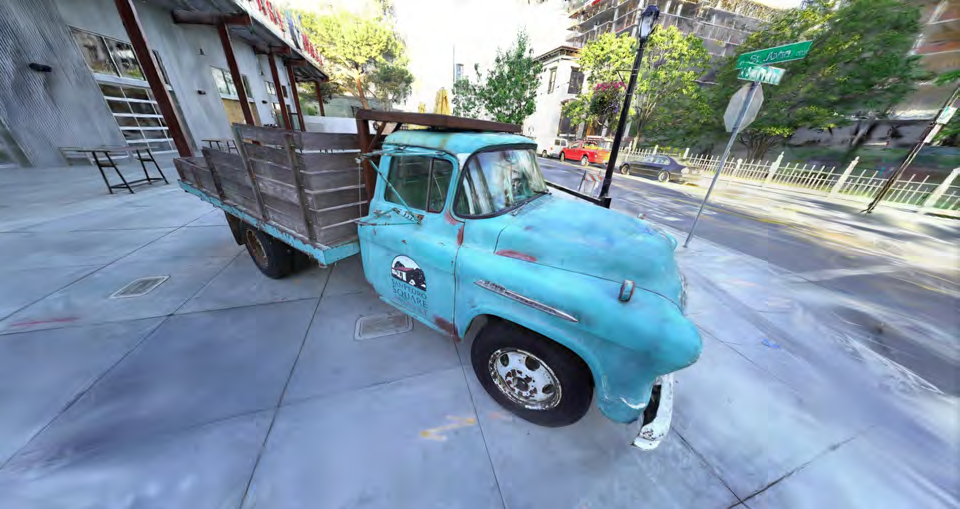}
    \caption{3D Gaussian Splatting scene of a truck. The scene is provided by the original paper and imported and rendered in our Vulkan renderer.}
    \label{fig:og_truck}
    \postcapspace
\end{figure}

In \gls{nvs}, a set of pictures of a scene is taken and used to create new images. These images may be generated from arbitrary viewpoints, beyond those included in the original set of pictures. The objective is to make these synthetic images convincing; seemingly taken in the original scene. 
One popular method is \gls{3dgs}. In \gls{3dgs}, Gaussians are modified until virtual, rendered images match real pictures closely. These Gaussians are powerful descriptors that can accurately represent even complex real-world environments. \autoref{fig:og_truck} depicts such a novel view of a reconstructed scene using \Gls{3dgs}. 

% Terms and scope of the topic

\gls{3dgs} uses neither mesh nor texture. Its only primitive is a 3D Gaussian, while the collectivity of 3D Gaussians represents a scene accessible in \gls{cpu} memory. As the scales of \gls{3dgs} scenes increase, so does the memory and work required to store and render them. Depending on the actual size of the scene, the amounts of data can be huge, multiple gigabytes or terabytes even. Handling \gls{3dgs} reconstructions requires graphics engines with high memory bandwidth along with huge amounts of memory, which seriously limits the application of \gls{3dgs} and puts its use on even modern mobile computers out of reach. Therefore, novel ideas and the invention of new methods to tackle these problems are of high relevance.

The method introduced in this work is concerned with these aspects of efficient data handling and rendering of large scenes.  We propose the use of virtual memory and \gls{lod} which help to achieve virtually unlimited scene scale and efficient rendering, even on low-end mobile devices. At a glance, virtual memory is used 
to handle such amounts of data and to provide the means for efficient rendering. Gaussians that are visible from a certain viewpoint are determined as quickly as possible and then transferred to \gls{gpu} memory. Thereby, Gaussians outside the view frustum or those occluded by other structures can be discarded and a significant amount of computation and memory can be saved. Through virtual textures that are continuously updated and exchanged between \gls{cpu} and \gls{gpu} memory, rendering large scenes, such as in open-world video games, becomes possible, even if the total size of the textures is larger than the available \gls{gpu} memory. Our approach is split into two parts: a \textbf{preprocessing} and a \textbf{real-time rendering} part. 

In the preprocessing stage, an existing \gls{3dgs} scene is analyzed offline to create a mesh that approximates the structure present in the scene. Gaussians are separated into pages that group them by their proximity. Each face on the proxy mesh is marked with the corresponding page ID. Where pages overlap, links between them are established. Finally, lower \gls{lod} levels for each page are provided.

% \begin{figure}[t]
% 	\centering
% 	\includegraphics[width=\linewidth,trim=0 20 0 20,clip]{05/demo/vbuffer}
% 	\caption{Image of a mesh approximating a \gls{3dgs} scene. Triangle primitives are shaded with an ID (in different shades) indicating which pages of Gaussians are likely visible. Since the shading involves drawing raw IDs without any lighting, there are little if any clues about depth. The geometry of the pictured scene is therefore nigh impossible to gauge.}
% 	\label{fig:intro_vbuffer}
% \end{figure}

The second part of our method is concerned with rendering scenes in real time. The preprocessed scene is imported and for each frame, we first render the page IDs on the proxy mesh to a visibility buffer. The sole purpose of this buffer is to find out the IDs of pages in view of the camera. The page IDs contained in the image, as well as their established links, allow us to determine which pages of Gaussians will likely be visible. The pages are then transferred to \gls{gpu} memory and rendered using a modern rendering API.

Note that we do not modify the process of scene creation. Any scene in the format described by Kerbl \etal~\cite{3dgs} is compatible with the approach presented. This includes the original implementation, similar derivative implementations, and at least a subset of more recent implementations that focused on increasing the size of reconstructed scenes. Several solutions have been proposed to reduce the memory usage or modify the attributes of \gls{3dgs} scenes. Many of these modifications may generally be used in combination with our work in the future. However, we limit our exploration of such extensions to a discussion of \gls{3dgs} compression in related work. 

% Outline the methodology
In the following, we provide details about the concepts involved, from preprocessing to per-frame operations. We explore several alterations to the basic idea to improve quality and performance or limit memory usage and highlight the resulting trade-offs. An end-to-end implementation is provided, together with an in-depth evaluation. Our contributions can thus be summarized as follows:
\begin{itemize}[leftmargin=*]
    \setlength\itemsep{0.0em}
    \item a novel approach to virtual memory and \gls{lod} within \gls{3dgs};
    \item an implementation for pre-processing and real-time rendering using a modern rendering API;
    \item an efficient method for rendering a set of large scenes on mobiles; and
    \item a detailed evaluation of our approach on desktop and mobile devices.
\end{itemize}

\section{Related Work}
\label{chap:related}
In this section we focus on the related work in \gls{3dgs}, which is relevant to our approach. Due to the vast amount of continuously published work, compiling a comprehensive overview is impossible. We also abstain from discussing the mathematical basics of \gls{3dgs} for brevity and refer the interested reader to the work of Kerbl~\etal~\cite{3dgs} for details. Because we introduce virtual memory to \gls{3dgs} in our work, we first briefly introduce the basics and then focus on related work required to understand the concepts. Thus, we bridge the gaps between \gls{3dgs} and virtual texturing.

\subsection{3D Gaussian Splatting}
%Image-based rendering~\cite{ibr}, or novel view synthesis, aims at rendering scenes, typically captured in the real world, entirely based on images. %In photogrammetry, scenes are captured and reconstructed, resulting in a textured triangle mesh, which can be rendered with a traditional rendering pipeline. Image-based rendering, in comparison, often circumvents these pipelines entirely. 
%Methods, such as light fields and \gls{nerf}, do not attempt to model the geometry of the scene. Although the novel views generated can be incredibly accurate, these representations present challenges in editing and interacting with scenes.
%
In \gls{nvs}, research is divided mainly into two state-of-the-art approaches: \gls{nerf} and \gls{3dgs}. While there are similarities, the latter was introduced by Kerbl~\etal~\cite{3dgs} to render radiance fields, representing scenes using an explicit scene representation. 3D Gaussians, stored as their ellipsoid equivalents, are defined by position, rotation, scale, opacity, and view-dependent color in the form of spherical harmonics. This representation allows them to produce accurate images that can be rendered in real-time on desktop hardware and edited easily. Images are rendered with a tiled software renderer implemented in \gls{cuda}.

Taking inspiration from the initial proposal of \gls{3dgs}, researchers have explored a variety of further avenues and details, some of which are relevant to the concept proposed in this work.

\subsubsection{Mesh Extraction}
\label{sec:related_mesh_extraction}
% Overview (marching cubes, sugar)
There are multiple reasons to create a triangle mesh from \gls{3dgs} scenes. The mesh may replace the scene representation entirely, or triangle meshes may support an application such as providing a foundation for physics interactions. 
The conventional method to extract a triangle mesh from a \gls{3dgs} reconstruction is to apply the Marching Cubes algorithm, proposed by Lorensen and Klein~\cite{marching_cubes}. The algorithm converts 3D data to triangle meshes depicting constant-density surfaces. % They acquire data with medical imaging machines, which combine multiple 2D images. 

%However, multiple papers point out that the Gaussians produced by \gls{3dgs} do not adhere to surfaces well enough to faithfully reconstruct smooth surfaces.
Chen \etal~\cite{neusg} introduce a regularization term to \gls{3dgs}, encouraging thin Gaussians to better align with surfaces. Gaussians, as well as a neural network learning the scene's \gls{sdf}, are jointly optimized to do this. Guédon \etal~\cite{sugar} introduce a similar regularization term. 
%They determine a level set by rendering a depth map of the scene. The rendered pixels are sample points that can be used to perform Poisson reconstruction before simplifying the resulting triangle mesh. 
Waczyńska \etal~\cite{games} use a parameterization that defines Gaussians by their position on a mesh. Both the mesh and its Gaussians are optimized together. The idea of a joint representation for mesh and Gaussians is especially promising for deformation and has also been explored in that context by Yuan \etal~\cite{gavatar} and Gao \etal~\cite{mesh_deformation}.

We don't aim to advance the field in this direction, rather we resort to using the Marching Cubes algorithm as a baseline for our work to also create a mesh for our purposes.

%We do not attempt to advance the field of mesh extraction. For use as a proxy mesh, we value limiting the number of primitives in the scene more than mesh quality. The proxy mesh needs only capture significant structures without exact details. We therefore fall back to the Marching Cubes algorithm, though the mentioned regularization terms (Chen \etal~\cite{neusg}, Guédon \etal~\cite{sugar}) can both improve mesh quality and make it easier to simplify.

% \begin{figure}[t]
% 	\centering
% 	\includegraphics[width=\linewidth,trim=0 18 0 0, clip]{05/scenes/scene_berlin}
% 	\caption{Image rendered using \gls{3dgs} with virtual memory. The indoor scene features many occlusions, making occlusion culling effective at improving memory use and rendering performance.}
% 	\label{fig:intro_vmem_example}
% \end{figure}

\subsubsection{Compression}

The concept of \gls{3dgs} implies the storage and handling of large amounts of data that increases significantly with the reconstruction scale. 
Kerbl \etal~\cite{3dgs} use 248 bytes to represent a single Gaussian. The majority of this storage is consumed by the spherical harmonics that depict colors. %As a result, even relatively straightforward scenes can include millions of splats. %Twelve bytes store the average color (the first spherical harmonics band), while 180 bytes encode additional view-dependent changes in color. Even a relatively simple scene can contain millions of splats. 

Pranckevičius~\cite{compr_arasp_1} creates groups of Gaussians by their positions, whose properties he stores in patches in 2D textures. %Since the properties within the groups are generally similar, the resulting textures can be compressed effectively with standard image compression algorithms. 
Morgenstern \etal~\cite{compr_compact3dscene} present an algorithm to efficiently sort millions of Gaussians by their properties in seconds to create 2D textures. 
%Pranckevičius~\cite{compr_arasp_1} additionally notes that reordering Gaussians, such as when using Morton order, can improve rendering performance simply by improving cache locality. We observe this effect in our work when grouping Gaussians into pages.
Later, Pranckevičius~\cite{compr_arasp_2} proposes using vector quantization with k-Means on spherical harmonics. Fan \etal~\cite{compr_lightgaussian} determine the importance of a Gaussian in a scene and use vector quantization on the less important Gaussians. Niedermayr \etal~\cite{compr_compressed_3dgs} determine the sensitivity of images to changes in properties to perform sensitivity-aware quantization. Vector quantization generates codebooks with an index to the respective entry for each Gaussian. Navaneet \etal~\cite{compr_compact3d} uses a form of \gls{rle} on these indices, while Niedermayr \etal~\cite{compr_compressed_3dgs} compresses them with the DEFLATE algorithm, which also employs a variation of \gls{rle}. 

%Other works train \gls{mlp} networks to reduce the size of their properties. 
Girish \etal~\cite{compr_eagles} quantize Gaussian properties, then decode those with a \gls{mlp} to recover the initial properties. Li \etal~\cite{compr_spacetime} train an \gls{mlp} to return a final color in dynamic scenes based on a base color, a view direction, and time. Lee \etal~\cite{compr_compact3dgau} determine a feature and a view direction from positions with a hash grid and use those in an \gls{mlp} to determine the color.

We do not employ any such compression in general, but our approach can be used alongside several of those extensions.

%In contrast to related works, we avoid compression in order to put additional stress on our system. Large scene reconstruction is still an active research topic and such tools are not readily available yet. Not compressing Gaussians allows us to test our method at its limits to effectively find shortcomings.

\subsubsection{Large Scene Reconstruction}
%Reconstructing large scenes has been a significant challenge to \gls{nvs} in recent years. With photogrammetry approaches, creating meshes from environments, classical computer graphics optimizations could be used. However, these mesh-based reconstructions were unable to recreate scenes accurately, especially when involving thin structures or transparencies. The introduction of \gls{nerf} allowed much more accurate reconstructions but their implicit nature makes them extremely difficult to scale up. 
\gls{3dgs} gains most of its popularity through its scalability to large scenes and the visual appeal of rendered imagery, albeit Kerbl \etal~\cite{hierarchical_3dgs}, complain about the lack of available huge datasets for further enhancements. Certainly, increasing the scale of reconstructions introduces a new range of challenges related to the amount of data that can be adequately stored in physical memory, making it essential to partition the reconstruction at some stage.

%\gls{3dgs}represents the next step in \gls{nvs} and shows promise in scaling to large scenes. Additionally, current research, like Kerbl \etal~\cite{hierarchical_3dgs}, complain about the lack of available datasets for such scenes. %With \gls{3dgs} lending themselves better to large scenes than its alternatives, more such datasets will likely be published.

%Expanding \gls{3dgs} involves modifying both reconstruction and rendering. The original implementation for scene creation requires large amounts of \gls{gpu} memory. All Gaussians, along with additional overhead, need to be available at all times. Current works attempting to increase scene size, all follow a similar approach. Inspired by attempts at scaling \gls{nerf}, scenes are separated into chunks, which are reconstructed separately and subsequently merged. This allows them to limit memory usage, only keeping the data relevant to a single chunk, and scale computation horizontally by reconstructing chunks in parallel.

A dataset and an approach to partitioning large scenes is proposed by Lin \etal~\cite{vast_gaussian}, splitting a scene into chunks based on the distribution of camera positions and the respective \gls{sfm} reconstruction. Liu \etal~\cite{city_gaussian} split scenes into uniform chunks and prune chunks with low contribution to rendering. Kerbl \etal~\cite{hierarchical_3dgs} define a low-resolution skybox on a sphere surrounding each chunk, defining the surrounding scene as background. Zhao \etal~\cite{scaling_up_3dgs} follows a different approach, distributing Gaussians across multiple \gls{gpu}s, assigning pixels to be rendered to \gls{gpu}s in tiles and transferring the Gaussians between them as needed. Dynamic load balancing can make distributed training efficient and scale to up to 32 \gls{gpu}s.

%Lin \etal~\cite{vast_gaussian} publish a dataset, containing large scenes reconstructed from aerial images in the original file format. We use a selection of these scenes for evaluation in Section~\ref{chap:evaluation}. 
Both Liu \etal~\cite{city_gaussian} and Kerbl \etal~\cite{hierarchical_3dgs} make their code for scene reconstruction available. However, the scenes created by Liu \etal~\cite{city_gaussian} are not available, and Kerbl \etal~\cite{hierarchical_3dgs} introduce a new file format to incorporate their \gls{lod} implementation. 
We use a selection of the scenes from Lin \etal~\cite{vast_gaussian} for evaluation in Section~\ref{chap:evaluation}. However, because of the differences in formats, direct comparisons to other implementations are generally challenging.

\subsubsection{Streaming}
\label{sec:related_streaming}

In their initial work, Kerbl \etal~\cite{3dgs}, only perform frustum culling during rendering. A tiled, compute-based software rasterizer determines whether a Gaussian overlaps with a tile before sorting Gaussians for each tile individually. To do this, the data for Gaussians need to be in \gls{gpu} memory.
%Frustum culling can be extended to stream Gaussians on demand with relatively little effort. 
Jiang \etal~\cite{3dgs_reloc} divide a scene into voxels. They determine which voxels lie in the view frustum using a KD-Tree and subsequently transfer these to the \gls{gpu}. Kerbl \etal~\cite{hierarchical_3dgs} asynchronously transfer high-detail Gaussians to replace those with lower detail as part of their \gls{lod} solution.

The use of streaming, or on-demand transfers, is still somewhat limited in current research. If it is used, the decisions for which Gaussians to copy is generally made based on the view-frustum, instead of any more sophisticated visibility determinations. In contrast, we perform visibility determination using a visibility buffer. Culling based on both the view frustum and occlusions are the result.

\subsubsection{Level of Detail}
\label{sec:related_lod}
Reducing the overall number of Gaussians to be rendered can significantly boost performance, % In return, visual quality suffers from reducing the number of Gaussians. With \gls{lod}, this tradeoff can be made based on the distance of any particular Gaussian from the camera. There are a few different approaches to implemented \gls{lod} in current research.
while the most common approach is to merge multiple Gaussians. Yan \etal~\cite{anti_alias_lod} propose combining Gaussians to solve aliasing problems in \gls{3dgs}. Gaussians that are too small during rendering are culled and replaced with larger ones. Kerbl \etal~\cite{hierarchical_3dgs} introduce a hierarchy of Gaussians, merging them based on the contribution of each one on its parent. Their hierarchy allows them to choose which layer and therefore \gls{lod} level to render at runtime.

Liu \etal~\cite{city_gaussian} do not reduce the number of Gaussians but compress those further away from the camera. Vector quantization proposed by Fan \etal~\cite{compr_lightgaussian} reduces the visual fidelity of less important Gaussians.
Lu \etal~\cite{scaffold_gs} create anchors in a scene, spawning neural Gaussians around each that can be dynamically adapted to the distance. Ren \etal~\cite{octree_gs} create an octree for a scene, where each level represents an \gls{lod} level, each containing anchors spawning additional neural Gaussians.

We base our \gls{lod} solution on Yan \etal~\cite{anti_alias_lod}. However, unlike most of the described methods, we do not modify Gaussians during scene reconstruction.

% VastGaussian - Training large scenes but slow to render
% CityGaussian - Additionally introduces lod
% 3DGS-Reloc - Splits into 2D voxels to stream to GPU, no occlusion culling
% https://repo-sam.inria.fr/fungraph/hierarchical-3d-gaussians/ - Hierarchical lod

\subsection{Virtual Texturing}

Virtual texturing is based on virtual memory, which modern operating systems and CPU architectures implement. Virtual memory separates virtual and physical address spaces. Memory addresses used by a process in userspace do not hint at actual locations in system memory but must be translated first. These translations query a page table, which maps pages of addresses in virtual memory to respective regions in physical memory once a process uses them. These pages are in regions of equal size in memory. The same system is used in modern \gls{gpu} architectures. Applications can create buffers or textures that are not bound to any physical memory initially. Regions of these can be manually bound once they are required. This functionality is exposed to applications through modern APIs as sparse residency (Vulkan) or reserved resources (Direct3D 12).

% http://vulkan.gpuinfo.org/listfeaturescore10.php - sparseResidencyBuffer
% https://www.asawicki.info/articles/memory_management_vulkan_direct3d_12.php
% https://www.asawicki.info/news_1698_vulkan_sparse_binding_-_a_quick_overview.html
% http://3dgep.blogspot.com/2016/02/a-journey-through-directx-12-dynamic.html

% https://silverspaceship.com/src/svt/
Barrett~\cite{sparse_virtual_textures} demonstrates how the concept can be applied to texturing, splitting textures into tiles (pages) of equal size. These tiles are generally not resident in \gls{gpu} memory if they are not visible. At runtime, the scene is first rendered with tile IDs only, and visible tiles are copied to the \gls{gpu} if necessary. An indirection texture points to the physical tile containing the required page. %Mip levels are precomputed and similarly split into tiles. At runtime, the appropriate level is selected. 
Mayer~\cite{vt_thesis} provides a detailed overview and analysis of virtual texturing methods. %They explore its history, the latest research at the time, and address problems.

Virtual texturing forms the foundation for our work. We use these ideas to determine which pages are visible and transfer them to the \gls{gpu} if necessary. Similarly to how mipmapping is integrated into virtual texturing, we integrate an \gls{lod} solution, and apply both concepts to improve \gls{3dgs}.

\section{Preprocessing for 3DGS using Virtual Memory}
\label{chap:vmem}
\begin{figure}[t]
	\centering
	\includegraphics[width=\linewidth]{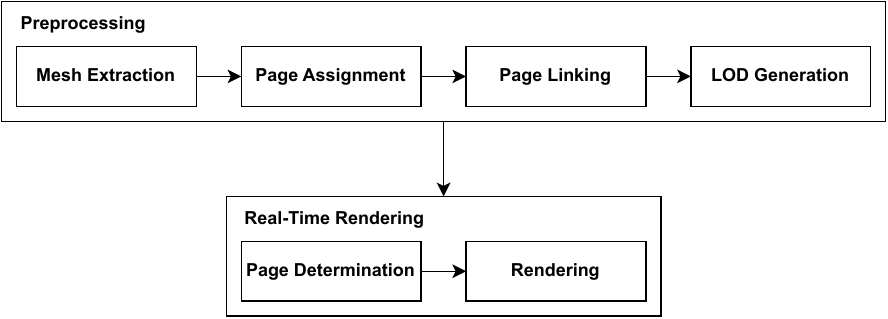}
	\caption{Scene pre-processing and real-time rendering steps: A preprocessing stage prepares the scene offline. Real-time rendering steps are performed for every rendered frame.}
	\label{fig:vmem_overview}
    \postcapspace
\end{figure}

Our virtual memory solution, inspired by virtual texturing, creates a mesh to approximate the \gls{3dgs} scene. Gaussians in the scene are grouped into pages. Rendering the proxy mesh quickly identifies which Gaussian pages are visible to the camera. In \autoref{fig:vmem_overview} we provide a visualization of the steps involved with the method, while this Section is concerned with the upper part only. The lower part of \autoref{fig:vmem_overview} is discussed in Section \ref{sec:implementation_realtime_rendering}. %\pf{our method has not been proposed yet}

% Preprocessing the scene prepares it for later real-time rendering. First we take a reconstructed \gls{3dgs} scene and extract a mesh. Gaussians are then assigned to fixed size pages. Each page is associated with an ID, which the proxy mesh is marked with. To properly detect page visibility, overlapping pages are linked. Finally, we create multiple \gls{lod} levels to further reduce memory requirements.

% First, the proxy mesh is rendered using the same viewport as the final scene. The image is analyzed to identify visible pages. Pages of Gaussians not in memory but needed for the render are transferred to the \gls{gpu}. Finally, visible Gaussians are sorted by distance from the camera and rendered.

\subsection{Mesh Extraction}

% concise
To render a scene proxy, we need a simplified mesh with fewer primitives for faster processing while maintaining enough faces to prevent page occlusion. We must avoid small holes from Gaussians to minimize unnecessary pages.
% To render a proxy of the scene we explicitly require a mesh with reduced complexity. We strive to render the triangle proxy mesh quickly, which is helped by reducing its primitive count. However, we require enough faces for a single page to lie adjacent. Details must be captured to avoid pages being occluded when they should not be. Simultaneously, small holes caused by Gaussians' lack of adherence to surfaces need to be avoided to reduce the number of pages required unnecessarily. We fall back on a solution based on Marching Cubes. \autoref{fig:vis_mesh_extraction} shows an overview of the steps involved.
Intersecting ellipsoids (3D Gaussians) with the image plane creates scene slices, similar to the medical imaging applications of the Marching Cubes algorithm. Alternatively, this data can be represented by a uniform grid of sample points, each potentially inside one or more ellipsoids. These sample points correspond to pixels in the rendered scene slices.

% Slice rendering + math
%We can intersect ellipsoids, which represent 3D Gaussians, with the image plane to create slices of a scene. This is the virtual equivalent of the medical imaging machines which the Marching Cubes algorithm was introduced to work with. An alternative interpretation of the same data is described by a uniform grid of sample points. Each sample point may be inside one or more ellipsoids in the scene. The sample points correspond to pixels in the rendered scene slices.

% \begin{figure}[t]
%     \centering
%     \includegraphics[width=\linewidth]{03/mesh_extraction}
%     \caption{Visualization of the steps involved in creating a triangular proxy mesh from a \gls{3dgs} scene. Slices are intersected with the scene to create images first. Morphological operations clean the data before creating a mesh using Marching Cubes.}
%     \label{fig:vis_mesh_extraction}
% \end{figure}
\begin{figure}[t]
	\centering
	\includegraphics[width=\linewidth,trim=0 40 0 40,clip]{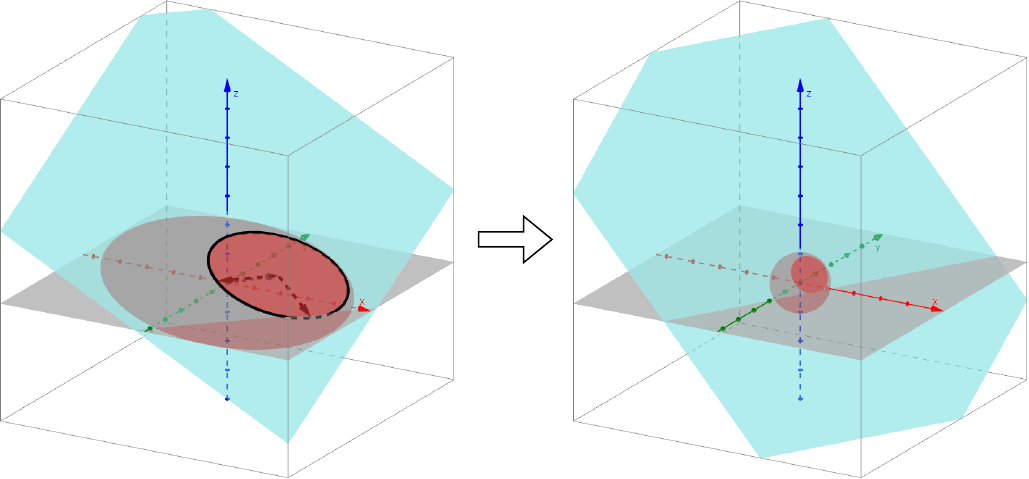}
  	\caption{The image depicts a red ellipsoid with its blue intersection plane, and the transformed unit sphere with the transformed plane on the right. When intersected with the image plane, ellipsoids form an ellipse (left). The ellipsoid is translated, rotated, and scaled into a unit sphere at the origin, with the plane adjusted accordingly (right). After determining the intersection circle, the original ellipse is recovered using inverse transformations.}
	\label{fig:ellipsoid}
    \postcapspace
\end{figure}

% Ellipsoid/plane intersection
To find the intersection between a plane and an ellipsoid, we use Hartmann's method~\cite{ellipsoid_math}. We define a plane normal to the Z-axis, adjusting the Z-coordinate as we scan the scene. Using a view matrix, we reposition the plane~$z=0$. The plane is in Hessian Normal Form, with its normal vector~$n=\begin{pmatrix} 0 & 0 & 1 \end{pmatrix}^T$, pointing in the positive Z direction, and its distance from the origin~$p=0$. We apply the ellipsoid's inverse transformations to simplify finding the intersection, resulting in a unit sphere at the origin. We adjust the plane by multiplying its normal vector by the Gaussian's inverse rotation matrix and normalizing it, as shown in Figure~\ref{fig:ellipsoid}. The distance from the origin is then determined by
% To find the intersection between plane and ellipsoid, we follow the method outlined by Hartmann~\cite{ellipsoid_math}. We define an intersection plane that is normal to the Z-axis. The Z-coordinate varies and changes as we sweep through the scene. Using a view matrix, the Gaussians' positions are modified to move the intersection plane to $z=0$. The plane can be expressed in Hessian Normal Form using the plane's normal vector ($n=\begin{pmatrix} 0 & 0 & 1 \end{pmatrix}^T$), in positive Z direction, and distance from the origin ($p=0$). We apply the ellipsoid's inverse translation, rotation, and scale to make finding the intersection easier. The result is a unit sphere at the origin. We compensate by modifying our intersection plane accordingly. Its normal vector is multiplied by the inverse of the Gaussian's rotation matrix and normalized. This process is pictured in Figure~\ref{fig:ellipsoid}. The distance from the origin is now determined by 
\begin{equation}
	p = \frac{v_z}{\|sn\|_2}
\end{equation} 
where $v$ represents the ellipsoid's translation in view space, and $s$ its scale. A plane intersects the sphere if the origin is closer than the unit sphere's radius, forming a circle centered at $c_0=np$ with radius $r=\sqrt{1 - p^2}$.
% where $v$ is the ellipsoid's translation in view space, and $s$ is its scale. If the distance from the origin is less than the radius of the unit sphere, the plane and sphere intersect. The intersection forms a circle centered at $c_0=np$ with a radius of $r=\sqrt{1 - p^2}$. 

\begin{equation}
    c_1 = \left\{ 
  \begin{array}{ c l }
    \vector{p \\ 0 \\ 0} & \quad \textrm{if } n_z = 1 \\\\
    \frac{p}{||sn||} \vector{(sn)_y \\ -(sn)_x \\ 0} & \quad \textrm{otherwise}
  \end{array}
\right.
    \label{eq:c1}
\end{equation}
\begin{equation}
    c_2 = \left\{ 
  \begin{array}{ c l }
    \vector{0 \\ p \\ 0} & \quad \textrm{if } n_z = 1 \\\\
    m \times c_1 & \quad \textrm{otherwise}
  \end{array}
\right.
    \label{eq:c2}
\end{equation}
We determine the parameters $c_1$ (Equation~\ref{eq:c1}), and $c_2$ (Equation~\ref{eq:c2}) to represent the circle as 
\begin{equation}
	c = c_0 + c_1 \cos{t} + c_2 \sin{t} \qquad,
\end{equation}
%\pf{hier würd eine skizze helfen zb siehe hier \url{https://ieeexplore.ieee.org/stamp/stamp.jsp?arnumber=10458391} fig 6}
which is subsequently transformed into an intersection ellipse using the Gaussian scale, rotation, and position, such that it can be drawn using its implicit representation.

% https://en.wikipedia.org/wiki/Ellipsoid#Determining_the_ellipse_of_a_plane_section
% https://mathworld.wolfram.com/HessianNormalForm.html
% https://en.wikipedia.org/wiki/Ellipse#General_ellipse_2
% https://en.wikipedia.org/wiki/Conjugate_diameters
% https://en.wikipedia.org/wiki/Rytz%27s_construction

%\pf{check comments for citatiosn}
% Klein, Peter Paul. “On the Ellipsoid and Plane Intersection Equation.” Applied Mathematics-a Journal of Chinese Universities Series B 2012 (2012): 1634-1640.
% https://www.semanticscholar.org/paper/On-the-Ellipsoid-and-Plane-Intersection-Equation-Klein/e07e844408c60f31f5b01e78cd6fe350d4fa7d75

% Ferguson, C.C. Intersections of ellipsoids and planes of arbitrary orientation and position. Mathematical Geology 11, 329–336 (1979). https://doi.org/10.1007/BF01034997
% https://link.springer.com/article/10.1007/BF01034997#citeas

% https://www.indianjournals.com/ijor.aspx?target=ijor:ijreas&volume=6&issue=6&article=027&type=pdf

% https://link.springer.com/article/10.1007/s00607-003-0060-0

Before applying Marching Cubes, we clean the data using morphological operations to close small holes and remove unnecessary detail, enhancing mesh simplification. These operations and kernels are tailored per scene for effective detail reduction. A surface is then reconstructed with Marching Cubes, and the mesh is simplified using a quadric error metric.

% Morphological operations
%Before using Marching Cubes, we clean the acquired data. Morphological operations close small holes and remove unnecessary detail. This improves the results of later mesh simplification stages. The specific operations and kernels must be adjusted for each scene to achieve the desired detail reduction. Finally, a surface is reconstructed with Marching Cubes, and the mesh is further simplified using a quadric error metric.

\subsection{Page Assignment}

% \begin{figure}[t]
%     \centering
%     \includegraphics[width=\linewidth]{03/page_assignment}
%     \caption{Visualization of the steps involved in assigning Gaussians to pages. The initial page assignment is based on which face of the proxy mesh a Gaussian is closest to. Subsequently, nearby pages are merged to reach a target page size. Based on their assignment, Gaussians are sorted, and small pages are padded.}
%     \label{fig:vis_page_assignment}
% \end{figure}

% Initial page assignment
% To avoid the overhead of determining each Gaussian's visibility separately, we group them into pages, using the steps shown in Figure~\ref{fig:vis_page_assignment}. Each face on the proxy mesh may correspond with either exactly one or no pages, though a page may span multiple faces. Each page has Gaussians assigned to it, with a defined maximum page size.

To minimize the overhead of checking Gaussian visibility, we group them into pages. A proxy mesh face can correspond to none or multiple pages, while a page can cover multiple faces. Each page holds Gaussians up to a specified maximum. Initially, each mesh face gets a separate page, and Gaussians are assigned to the page of the face nearest to their mean. If a face gets more Gaussians than the page size allows, it is subdivided by adding a vertex at an edge's midpoint, forming two triangles. This is repeated until each face has no more Gaussians than the target page size.

% To reduce the overhead of checking each Gaussian's visibility individually, we group them into pages. A proxy mesh face may correspond to one or no pages, while a page can span multiple faces. Each page contains Gaussians up to a set maximum size.
% To create this mapping given the constraints, we start by creating a separate page for each face of the proxy mesh. Gaussians are initially assigned to the page that corresponds to the face closest to its mean. 
% If this initial page assignment assigns more Gaussians to a single face than the page size allows, the face must be subdivided. A new vertex is inserted at the midpoint of one of its edges, creating two triangles in its place. This process is repeated until the number of Gaussians assigned to each face is no greater than the target page size.

% \begin{figure}[t]
%     \centering
%     \includegraphics[width=\linewidth]{03/edge_rotation}
%     \caption{We rotate the vertex order in the two triangles created by a subdivision, as demonstrated on the right. This avoids subdividing at the same edge multiple times, which can lead to the creation of thin triangles (left).}
%     \label{fig:edge_rotation}
% \end{figure}

To prevent thin triangles from repeated subdivisions, we rotate the vertex order and choose different edges for subsequent subdivisions. The method isn't optimal because shared edges may create vertices on adjacent triangles, leading to T-junctions and rendering artifacts due to floating point inaccuracies.

% To avoid repeated subdivisions creating thin triangles, we rotate the order in which the vertices appear for the two triangles. A different edge is chosen if a face created by one subdivision requires another subdivision. The difference is visualized in Figure~\ref{fig:edge_rotation}. 
% The described method to subdivide a triangle is not optimal. Since the edge of any triangle is likely shared between two triangles, subdividing only one of the two creates a vertex on another triangle's edge. Floating point inaccuracies lead such T-junctions to result in minor rendering artifacts.

% Merging
Initial page assignments group nearby Gaussians but create small pages. We merge neighboring pages to approach the target size without exceeding it, using a greedy, breadth-first method. Pages are first merged with neighboring faces. If a page remains under the target size and no neighbors are left, a nearby page is chosen. If adding a page exceeds the target size, it becomes the base for a new merged page, to which more pages are added.

% The resulting initial page assignment groups nearby Gaussians but generally creates small pages. We merge neighboring pages to increase page sizes, approaching but never exceeding the target. To do so, we add pages in a greedy, breadth-first manner. Pages are first merged with their respective faces' neighbors. If a page has not reached the target size and no more neighbors are available, another page is chosen by proximity. Since we merge pages greedily, we may attempt to add a page whose size would increase the merged page's size beyond the target. Once this happens, this last page will become the base for a new merged page, to which nearby pages will be added.

The initial pages, each per face, are greatly reduced. A triangular mesh face maps to one or no page, marked with a page ID for rendering if included. Unmarked faces lack a Gaussian mean match during initial assignment. Gaussians are grouped in memory once assigned to pages, with zero-property Gaussians added as padding to reach page size, easily culled during rendering.

% As a result, the number of initial pages, one per face, is significantly reduced. A face on the triangular proxy mesh maps to either one or no page. If a face belongs to a page, it is marked with its page ID for page determination during rendering. The face remains unmarked if no Gaussian mean was closest to a face during the initial page assignment.
% Once all Gaussians have been assigned to a page, they can be grouped in memory accordingly. Additional Gaussians, with all properties set to zero, are added as padding to meet the target page size. This padding can easily be identified and culled while rendering.

\subsection{Page Linking}
\label{sec:vmem_prep_links}

% \begin{figure}[t]
%     \centering
%     \includegraphics[width=\linewidth]{03/page_linking}
%     \caption{Visualization of the steps in creating links between pages where overlaps occur. Sample points within each ellipsoid are generated, whose closest face on the mesh is found. Links are created based on its assigned page, and unassigned faces are assigned a page.}
%     \label{fig:vis_page_linking}
% \end{figure}

\begin{figure}[t]
    \centering
    \includegraphics[width=1\linewidth]{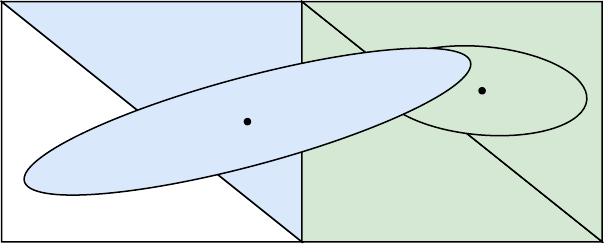}
    \caption{2D demonstration of a Gaussian, which has been assigned to a page based on its position, overlapping with another nearby page. Without page linking, the visibility determination for such a Gaussian may be incorrect.}
    \label{fig:linking}
    \postcapspace
\end{figure}

% Problems with page assignment
In \gls{3dgs} scenes, millions of overlapping ellipsoids don't align with surfaces like textures do with mesh faces. We assign Gaussians to pages based on their means, without considering nearby interactions. Figure~\ref{fig:linking} shows how ellipsoids extending into surrounding pages can incorrectly be marked as invisible.
% Unlike textures, which map directly onto mesh faces, \gls{3dgs} scenes contain millions of ellipsoids. They inevitably overlap and do not align precisely with surfaces. We use Gaussians' means when assigning Gaussians to pages but disregard how they interact with other nearby pages. Figure~\ref{fig:linking} demonstrates how ellipsoids reach into surrounding pages, which can cause them to be mistakenly marked as invisible.

We address the issue by using page linking, identifying overlapping pages after initial assignment to establish uni- or bi-directional links. When a page is visible, all linked pages must also be present in \gls{gpu} memory.
% To overcome this problem, we introduce the concept of page linking. After creating an initial page assignment, we can determine which other pages a Gaussian overlaps with to create a link between the respective pages.
% %(Figure~\ref{fig:vis_page_linking}). 
% Such a link can be established uni- or bi-directionally. Any linked page is also required whenever a page is marked as visible. Both pages, therefore, need to be present in \gls{gpu} memory.
% Algorithm
To naïvely find pages that need linking, each ellipsoid is tested for intersections with others. Intersecting ellipsoids on different pages require links. Lacking acceleration structures, this approach has $O(n^2)$ time complexity, leading to trillions of tests even for small scenes.
% To determine pages that require linking naively, all ellipsoids would be tested for intersections with all other ellipsoids. Intersecting ellipsoids assigned to separate pages indicate the need for a link. Without acceleration structures, such an algorithm has $O(n^2)$ time complexity, resulting in trillions of intersection tests, even for small scenes.
Instead, we generate random points within an ellipsoid and identify the nearest face on the proxy mesh for each. A link is needed if the face's page is different from the ellipsoid's assigned page. This approximate method tests only the closest face on the low-poly proxy mesh. This step is part of our algorithm for initial mesh assignment, allowing optimized page assignment and linking.
% Instead, we create randomly sampled points within an ellipsoid. We find the closest face on the proxy mesh for each sample. A link is required if the face's page differs from the page to which the ellipsoid is assigned. This method is an approximation; it only tests for the closest face on the low-poly proxy mesh. Additionally, determining the closest mesh face to a given point is already required in our algorithm for initial mesh assignment, enabling joint optimizations for page assignment and page linking.
Random points are generated uniformly in a unit sphere using spherical coordinates and converted to Cartesian coordinates. Applying Gaussian scale, rotation, and translation transforms these points to lie within an ellipsoid. The final distribution is non-uniform due to initial sampling in the unit sphere, not accounting for the ellipsoid's properties.
% We initially create uniformly distributed random points within a unit sphere in spherical polar coordinates. The samples are converted to the Cartesian coordinate system. Finally, we apply the Gaussian's scale, rotation, and translation to the samples. The resulting random samples lie within the ellipsoid. Note that the final sample distribution is not uniform because sampling is done in a unit sphere without accounting for the ellipsoid's scale and rotation.

% Link pages
After assigning Gaussians to pages, a face on the proxy mesh may match one page. If the closest face for a sample belongs to a different page than its Gaussian, a link is established.
% After Gaussians are assigned to pages, faces on the proxy mesh may correspond to one page. Once the closest face for a sample is found, we compare the page the face belongs to to the page the Gaussian is assigned to. If they are not the same, a link is created.
% New pages
If a mesh face isn't yet assigned to a page, it is marked with the page having the most sample points close to it. The page with the most overlap takes precedence, while other overlapping pages are linked. The number of page links indicates the quality of page assignments, which should aim to be compact and minimize overlaps. Future work may refine assignments based on these links.
% If the closest mesh face to a sample point does not correspond with a page yet, the face can be marked. We mark the face with the page with the most sample points closest to the new face. Therefore, the page that overlaps with a previously unmarked face the most takes precedence. All other pages with overlaps are then linked normally.
% The number of page links can give us some insight into the quality of how pages have been assigned. Good page assignments should strive to keep pages compact and avoid large overlaps between pages. While we do not take this into account, future work may choose to refine page assignments based on the number of page links.

\subsection{Level of Detail}
% Why
In mesh rendering, a mipmap is a texture with multiple precomputed, downscaled versions. As surfaces move further from the camera, lower resolution levels are used, improving performance by reducing texel fetches and offering anti-aliasing through bilinear filtering. Early virtual texturing, like Mittring \etal~\cite{advanced_vt}, relies on mipmaps. The mip level is chosen when transferring a new texture to the \gls{gpu}.
% In mesh rendering, a mipmap is a texture for which multiple downscaled options are precalculated. As the distance of surfaces from the camera increases, texels are sampled from a higher level and, therefore, a lower resolution image. Mipmapping is generally effective in improving performance due to the reduced texel fetches and anti-aliasing since textures are typically downscaled with bilinear filtering. Even early implementations of virtual texturing rely on mipmaps, as demonstrated by Mittring \etal~\cite{advanced_vt}. The mip level is specified when requesting to transfer a new texture to the \gls{gpu}. 

Mip level selection is performed based on derivatives. To calculate these derivatives, rendering is performed in groups of two by two pixels, known as quads. All pixels in such a quad are rendered for a primitive, even if the primitive only overlaps a single pixel. The pixels that do not overlap are helper pixels, executed in helper lanes, and are subsequently discarded. This can cause an overhead of up to three additional pixels per pixel rendered to the output image. 
%The effect is especially pronounced with very small primitives. The triangle primitive in Figure~\ref{fig:helper_pixels} contributes to four pixels in the final image. However, its size and position cause it to overlap with four separate quads. Sixteen pixels are rendered, therefore, of which twelve are discarded.
% \begin{figure}[t]
%     \centering
%     \includegraphics[height=4cm]{03/helper_pixels}
%     \caption{Example for how a small triangle can cause overhead during rendering. The triangle primitive overlaps four pixel-centers according to the top-left rule. All four pixels are part of a separate quad, causing three additional helper pixels to be calculated for each pixel contributing to the output.}
%     \label{fig:helper_pixels}
% \end{figure}
\gls{lod} offers a solution with a mesh having various complexity levels, switchable based on camera distance, minimizing small primitives and reducing helper lanes, thus lessening performance impact.

% \gls{lod} is one solution to this problem. It generally describes a mesh with multiple versions of different complexities. They can be interchanged at runtime, often based on their distance from the camera. This reduces the number of small primitives, reducing the number of helper lanes and, therefore, reducing their performance impact.

\begin{figure}[t]
    \centering
    \includegraphics[width=\linewidth]{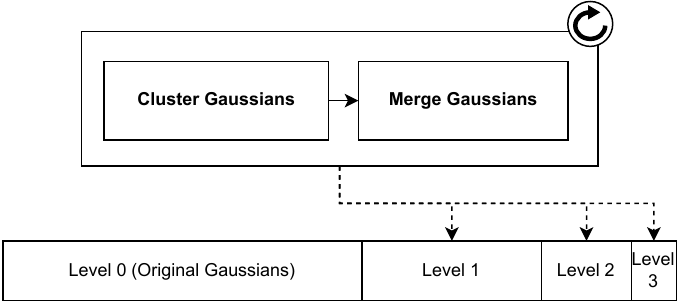}
    \caption{Visualization of \gls{lod} generation for four total levels (including the original Gaussians as level zero). The number of Gaussians per page is halved for each subsequent level. Merging Gaussians is done with an average for all properties but their scale.}
    \label{fig:vis_lod_generation}
\end{figure}

% Concept, reference paper
Yan \etal~\cite{anti_alias_lod} propose a method for \gls{3dgs} using mipmapping and \gls{lod}, merging Gaussians during training to create detail levels and reduce aliasing. We develop an \gls{lod} solution (Figure~\ref{fig:vis_lod_generation}) based on this, but focus on minimizing the Gaussians to be copied, stored in \gls{gpu} memory, sorted, and rendered, without altering the scene creation stage.

% Yan \etal~\cite{anti_alias_lod} propose a method for \gls{3dgs} inspired by mipmapping and \gls{lod}. They merge Gaussians during training, creating levels of detail to combat aliasing. Based on their approach, we create an \gls{lod} solution, as seen in Figure~\ref{fig:vis_lod_generation}. They do this to reduce aliasing, by filtering out Gaussians with low pixel coverage and replacing them with larger merged Gaussians. Unlike them, our goal is to reduce the number of Gaussians that need to be copied, stored in \gls{gpu} memory, sorted, and rendered. Additionally, we do not modify the scene creation stage.

\begin{figure}[t]
    \centering
    \includegraphics[width=\linewidth,trim=0 8 0 0,clip]{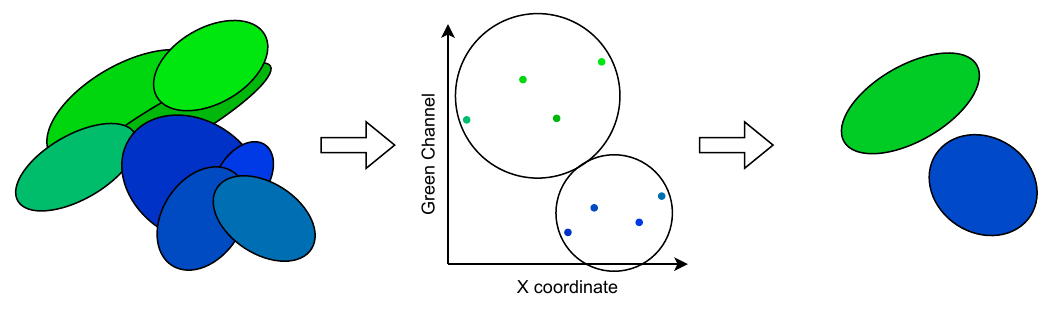}
    \caption{Gaussians are clustered based on their properties, then averaged. This illustration clusters ellipses based on only their x coordinates and green channels. The ellipses are then averaged such that each cluster becomes one new ellipse.}
    \label{fig:clustering_vis}
    \postcapspace
\end{figure}

% K-Means
The objective is to merge Gaussians while minimally impacting their distant appearance. Only Gaussians on the same page can be merged. Their attributes are points in multi-dimensional space, making the task one of clustering these points by similar attributes such as position, rotation, and color. By scaling attributes, their influence on clustering is adjusted. Clusters are identified using the k-means algorithm. Figure~\ref{fig:clustering_vis} illustrates the concept by clustering and averaging ellipses based on two properties.

% The goal is to merge Gaussians while keeping the impact on their appearance at a distance minimal. Only Gaussians within the same page can be merged to work with the rest of our solution. The Gaussians' attributes can be considered points in a multi-dimensional space. Thus, the problem is reduced to finding clusters of these points, which indicates Gaussians with similar attributes, such as position, rotation, color, etc. Attributes can additionally be scaled to change their impact on the resulting clustering. Clusters can then be found with the k-means algorithm. The concept is demonstrated simply in Figure~\ref{fig:clustering_vis}, clustering ellipses based on two of their properties, then averaging them.
% Scale problem
Merging multiple Gaussians is straightforward for most attributes using the arithmetic mean for translation, rotation, opacity, and spherical harmonics. However, as Yan \etal~\cite{anti_alias_lod} note, merging scales can cause the resulting Gaussian to be too small. To remedy this, additional Gaussians are adjusted using average pixel coverage during training. We scale up the merged Gaussian by a fixed factor to fill gaps. In clustering, we prioritize position attributes to prevent dispersed clusters.

\subsection{Scene Storage}
\label{sec:implementation_preprocessing_gaussians}
% % Loading
The original \gls{3dgs} implementation exports generated data in \gls{ply}. Other applications, such as Nerfstudio~\cite{nerfstudio}, have followed suit. The data describing such a scene consists of a list of Gaussians with 62 properties each.
% Decoding, transformations
The original encoding ensures quality despite numerical errors. Properties must be decoded from disk to display. Our virtual memory solution streams Gaussian to GPU memory when needed, possibly loading just in time. Thus, we eliminate encoding and adjust the file format to stream data directly to the GPU. These modifications to the original format include:
\begin{itemize}[leftmargin=*]
    \setlength\itemsep{0.0em}
    \item removing obsolete normal properties;
    \item reordering spherical harmonics data for better cache-locality;
    \item decoding the opacity property;
    \item decoding the scale properties;
    \item normalizing the quaternion describing the rotation.
\end{itemize}
We defer compression, such as converting spherical harmonics to half floats, to future work. Coordinate conversions, if needed, can be done by adjusting position, scale, and rotation.
% Culling
We limit the \gls{3dgs} scene to a central cube, excluding Gaussians with means outside it. %This isn't essential but simplifies slice rendering and mesh extraction. 

The mesh is stored alongside the modified scene file, including the following primitive attributes:
\begin{itemize}[leftmargin=*]
    \setlength\itemsep{0.0em}
    \item a list of page IDs, with each entry corresponding to a mesh face;
    \item a buffer containing page links;
    \item metadata such as page size, number of \gls{lod} levels, etc.
\end{itemize}

\section{Real-Time Rendering}
\label{sec:implementation_realtime_rendering}

Rendering is time-sensitive. Modern applications require at least 60 \gls{fps}, or frame times below a certain threshold, \eg $16$ ms. Virtual reality demands even higher rates to prevent nausea. Integrating rendering with other processes (e.g., physics simulations, pathfinding) can reduce the time available for each frame.
% Rendering is incredibly time sensitive. In a modern application, at least 60 \gls{fps}, that is a frame time below $\sim16$ms are generally expected. For virtual reality, even higher frame rates may be necessary to avoid undesired effects such as nausea. Additionally, if this technology is to be integrated into an application with other work (physics simulations, pathfinding, etc.), the time available to dedicate to rendering each frame may be shortened drastically.
Our real-time rendering concept allows for low-level optimizations. Parallelizable tasks are handled in thread pools or with compute shaders.

% Vulkan is a modern graphics API we use to interact with the \gls{gpu}, whether with graphics or compute tasks. Modern graphics APIs (Vulkan, Direct3D 12, Metal) distinguish themselves from their predecessors (OpenGL, Direct3D 11) with much greater explicit control over all aspects of \gls{gpu} configuration and execution.

%\subsection{Scene Loading}
% File format description
% The original \gls{3dgs} format is modified as described in Section~\ref{sec:implementation_preprocessing_gaussians}. These modifications simplify importing Gaussians. 
% All properties can be passed to the vertex shader as they are read from the file. Besides the Gaussians, we store the proxy triangle mesh in \gls{gltf}. The mesh primitive is accompanied by
% \begin{itemize}[leftmargin=*]
%     \setlength\itemsep{0.0em}
%     \item a list of page IDs, with each entry corresponding to a mesh face;
%     \item a buffer containing page links;
%     \item metadata such as page size, number of \gls{lod} levels, etc.
% \end{itemize}

% Memory mapping
The proxy mesh loads instantly, but Gaussians use a memory-mapped file placed in virtual memory, with physical memory accessed on a page fault. Although physical access is slower and generally undesirable, this method allows \gls{3dgs} scene size to exceed system memory limits.

% original:
% While the proxy mesh is loaded immediately, Gaussians are opened in a memory-mapped file. The entire file is placed in virtual memory. However, the operating system may only read from the file and put it in physical memory upon a page fault. While such an on-demand copy is slow and should be avoided, we are not limited in the size of the \gls{3dgs} scene by system memory as a result.

\subsection{Page Determination}

% \begin{figure*}[t]
%     \centering
%     \includegraphics[width=\linewidth]{03/page_determination}
%     \caption{Visualization of the steps in rendering Gaussians using virtual memory. These real-time rendering steps are performed for each frame. First, the proxy mesh is used to determine visible pages. Newly required pages are copied to the \gls{gpu}. Only visible pages are sorted by their distance from the camera. Finally, the scene is rendered.}
%     \label{fig:vis_page_determination}
% \end{figure*}

% Proxy mesh rendering
Before rendering a frame, we must determine which pages are visible from the camera. The aim is to cull Gaussians lying outside the view frustum and those occluded by other Gaussians. The Gaussians assigned to these visible pages must be copied if they are not yet in \gls{gpu} memory.

For virtual texturing, Barrett~\cite{sparse_virtual_textures} suggests rendering scenes with page IDs. All these IDs are needed for the final frame. We adopt this by marking our proxy mesh faces with page IDs, generating a list of visible pages from the image. Inspired by operating systems, we use a page table where each entry maps to a physical memory page, detailing virtual page properties. After listing required pages, new ones replace unused physical pages, prioritizing the least recently used.
% For virtual texturing, Barrett~\cite{sparse_virtual_textures} proposes solving this by rendering the scene with page IDs. All page IDs in the resulting image are required to render the final frame. We follow this approach by rendering our proxy mesh with its faces marked with corresponding page IDs. We can generate a list of visible pages by analyzing the resulting image. We maintain a page table inspired by virtual memory in operating systems. Each entry in the page table corresponds to a page in physical memory and stores properties about the virtual page stored within it. After creating a list of required pages, newly required pages can be copied into physical pages containing unused pages. The least recently used pages are prioritized when replacing pages in physical memory.
Once visible Gaussian pages are loaded into \gls{gpu} memory to render the scene, we include the sorting of Gaussians by camera distance for alpha blending.
% After determining which pages of Gaussians are visible and ensuring they are present in \gls{gpu} memory, the scene can be rendered. This includes sorting the individual Gaussians by their distance from the camera for alpha blending.
%Figure~\ref{fig:vis_page_determination} contains an overview of all the steps performed to render a frame.

% Page linking, lod
\paragraph{Page Links:}
We request pages via links, as explained in Section~\ref{sec:vmem_prep_links}, where we preprocess scenes to identify overlapping pages. Each page's visibility buffer includes its overlapping pages, and marking them as required minimizes visible artifacts.
% We can additionally request pages based on page links. In Section~\ref{sec:vmem_prep_links}, we describe how we preprocess a scene to determine which pages overlap. Each page in the visibility buffer is associated with a list of its overlapping pages. By traversing through the list and also marking these pages as required, we reduce visible artifacts.

\paragraph{Level of Detail:}
We add the distance of the nearest pixel to the visible pages list with its page ID. As the page table updates and new pages are copied, a suitable \gls{lod} is chosen based on this distance.
% We augment the list of visible pages with the distance of the closest pixel with the corresponding page ID. As the page table is updated and new pages are copied, an appropriate \gls{lod} is selected based on this distance.
This resembles choosing a mip level in virtual texturing. Barrett~\cite{sparse_virtual_textures} segments mip levels into uniform-sized pages. We instead reduce Gaussians with each \gls{lod}, allowing multiple virtual pages within one physical page. We mark page table entries with their current \gls{lod} for this reason. 
% This is akin to selecting a mip level to sample from in virtual texturing. Barrett~\cite{sparse_virtual_textures} segments all mip levels into pages of the same size. We take a different approach by reducing the number of Gaussians with each \gls{lod}. Multiple virtual pages may, therefore, be present within a single physical page. We mark a page table entry with its current \gls{lod} to allow for this.

Thresholds, which can be dynamically updated based on memory usage, move away from the camera when usage exceeds a limit and move closer when memory is available.
% The distance at which the \gls{lod} can easily be determined based on static thresholds. These thresholds may also be updated dynamically, increasing or reducing the number of Gaussians based on the ratio of used memory. If the ratio exceeds a set limit, the closest threshold is moved further away from the camera. All other thresholds can be set based on the first threshold, making them follow. When enough memory is available, and lower levels are still in use, the thresholds can be moved closer to the camera.
However, setting the initial threshold in fixed increments is problematic;
\begin{itemize}
    \item If the step size is too large, a single step may affect memory usage drastically, overshooting the targeted region. This can lead to rapid flipping between two levels, causing noticeable artifacts.
    \item Differences in scene scale can make a step size too large, leading to the above mentioned issue, or too small, making the system slow to adapt.
\end{itemize} 
Thus, the step size should be dynamically adjusted based on past moves to adapt plausibly.

\subsubsection{Page Table Management}
\begin{figure}[t]
    \centering
    \includegraphics[width=\linewidth,trim=0 2 0 0,clip]{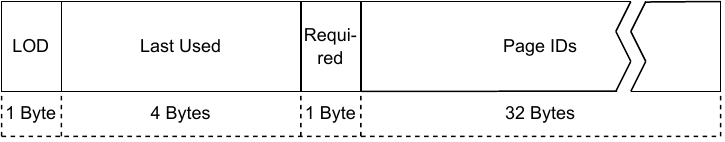}
    \caption{Visualization of the page table entry for a single physical page. The entry contains the page's current level of detail and the frame index for which the page was last required. For each virtual page within the physical page, a field indicates whether the page is required and the present page IDs.}
    \label{fig:pte_vis}
    \postcapspace
\end{figure}

To track pages in \gls{gpu} memory, we use a page table with entries describing the physical page contents, detail level, last use, necessity for the frame, and a detail-dependent list of contained pages.
%To keep track of pages in \gls{gpu} memory, we construct a page table consisting of a list of page table entries. Each entry describes the contents of a single physical page. It includes the level of detail, the page's last use, whether the page is required for the frame, and a list of all the pages contained, for which the maximum number is determined based on the level of detail. 
Figure~\ref{fig:pte_vis} shows this page table entry. We selected a single array structure over a hierarchical one for better memory locality. The page table's sequential iteration avoids field access indirection.
%Such a page table entry can be seen in Figure~\ref{fig:pte_vis}. We choose this structure, with all data contained in a single array, rather than a hierarchical structure for optimal memory locality. The page table is iterated sequentially, with none of the accesses to any of its fields leading to indirection.

Page table updating involves two steps: First, it is traversed to update which physical pages are required based on the list of required page IDs. Free physical pages that is, pages that contain no valid page or pages that are not necessary for the current frame, are stored along with the frame in which they were last used. The list of required pages is simultaneously updated, removing pages already present in the page table at the correct \gls{lod}.
%The page table is updated in two stages. First, it is traversed to update which physical pages are required based on the list of required page IDs. Free physical pages that is, pages that contain no valid page or pages that are not necessary for the current frame, are stored along with the frame in which they were last used. The list of required pages is simultaneously updated, removing pages already present in the page table at the correct \gls{lod}.

In stage two, the system checks the updated list of required pages, which contains page IDs to render the frame, none of which are in \gls{gpu} memory. If there's space, these pages are copied to a staging buffer, updating the page table. Free physical pages are chosen based on the least-recently-used order, then moved from the staging buffer to the Gaussian buffer on the \gls{gpu}.
%In the second stage, the updated list of required pages is traversed. It now contains page IDs required to render the frame, none of which are present in \gls{gpu} memory. If space remains, these pages are copied to a staging buffer, and the page table is updated accordingly. Free physical pages are selected in the least-recently-used order. The pages are subsequently copied from the staging buffer into the buffer of Gaussians on the \gls{gpu}.
%
We use a regular buffer, skipping the sparse residency feature described in Section~\ref{chap:related}, for three main reasons:
%We choose to use a regular buffer to do this, foregoing the sparse residency feature described in Section~\ref{chap:related}. There are three main reasons:
\begin{itemize}[leftmargin=*]
    \setlength\itemsep{0.0em}
    \item In virtual texturing, textures are addressed using UV coordinates, typically needing translation via an indirection texture, adding a lookup overhead. We avoid this lookup and use an index buffer to indicate Gaussians for drawing, which must be sorted regardless of sparse buffers.
    %In virtual texturing, textures are addressed using UV coordinates, which need to be translated first. This is usually done with an indirection texture, which adds the overhead of another texture lookup. We have no such lookup to be optimized. We signal the Gaussians requiring drawing using an index buffer, which needs to be sorted regardless of any sparse buffers.
    \item Only a few newer high-end desktop \gls{gpu}s support it. The hardware capability database~\cite{vulkan_gpuinfo} indicates that less than a third of \gls{gpu}s permit the \texttt{sparseResidencyBuffer} feature. For Android devices, support is just over 10\%. Without MoltenVK~\cite{moltenvk}, bridging Vulkan and Metal, Apple devices generally lack this feature.
    %Only a limited subset of hardware, typically more recent high-end desktop \gls{gpu}s, supports it. According to the hardware capability database~\cite{vulkan_gpuinfo} less than a third of \gls{gpu}s allow enabling the \lstinline|sparseResidencyBuffer| feature. For Android devices, this number is barely over 10\%. Without support in MoltenVK~\cite{moltenvk}, which forms a layer between Vulkan and Metal, this feature is generally not available on Apple devices.
    \item Even on recent hardware with sparse residency, performance seems to be unsatisfactory as mentioned by Richermoz~\cite{sparse_performance}.
\end{itemize}

% Index buffer construction
The list of physical pages is written to a buffer to indicate required pages to the renderer. The compute shader is adjusted to convert these pages into sortable indices based on camera distance.
\subsection{Rendering}
\begin{figure}[t]
	\centering
	\includegraphics[width=\linewidth]{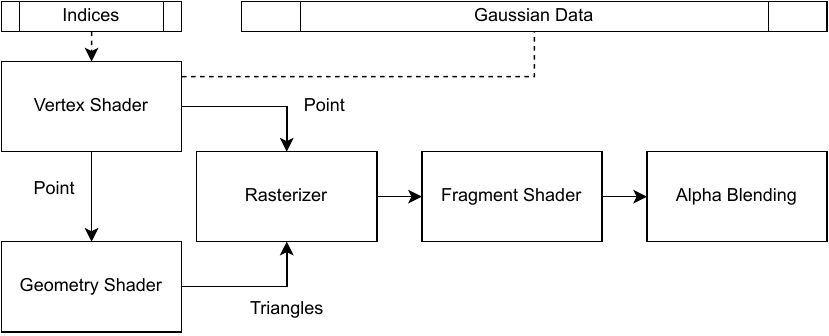}
	\caption{Simplified overview of our pipeline used to render Gaussians. A vertex shader is supplied with a sorted index buffer, which it uses to retrieve properties for its respective Gaussian from a storage buffer. If the device is capable of a geometry stage, a geometry shader is used to construct a rectangular quad. Alternatively, a point primitive is used. A fragment shader draws an ellipse onto the quad. Finally, the quads are combined via alpha blending to create the final image.}
	\label{fig:render_pipeline_overview}
    \postcapspace
\end{figure}

% Hardware rasterizer vs compute
There are generally two approaches to rendering \gls{3dgs} scenes. Kerbl \etal~\cite{3dgs} use CUDA to create a software renderer. The image is split into tiles, determining the Gaussians overlapping with each. Each tile's Gaussians are sorted and then drawn to the output image.
Our implementation instead opts to sort Gaussians globally in a \gls{glsl} compute shader before splatting the Gaussians onto quads rendered using the hardware rasterizer.

\begin{figure}[t]
    \centering
    \includegraphics[width=\linewidth,trim=0 5 0 5,clip]{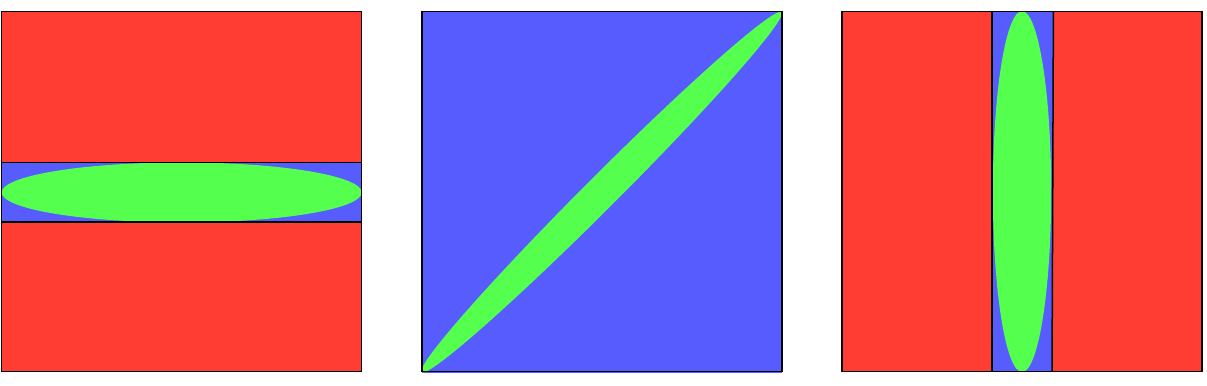}
    \caption{The area resulting in discarded pixels when drawing a long, thin ellipse. A point primitive is always square, resulting in fragment shader invocations that lead to a discard for pixels in the red and blue areas. Only the blue area is discarded when using a geometry shader to fit the quad more tightly to the ellipse. In the second image, the rotation of the ellipse causes any advantages to be lost, the overdraw is the same with and without a geometry shader.}
    \label{fig:point_overdraw}
\end{figure}

% Index buffer
The quads can be constructed in multiple ways, as depicted in Figure~\ref{fig:render_pipeline_overview}. One option is to record and submit commands to draw point primitives. This dispatches a vertex shader for each Gaussian, performing the necessary calculations. If geometry shaders are available, a geometry shader constructs the quad by emitting four vertices in a triangle strip. Finally, a fragment shader discards pixels outside the splatted ellipsoid.
If geometry shaders are unavailable, however, we can fall back to a pipeline, omitting the geometry stage. The quads are then the result of the point primitives, passed to the rasterizer. Their square shape can lead to a significant increase in fragment shader invocations, leading to discards. In Figure~\ref{fig:point_overdraw} three extreme cases are explored. The area shaded red indicates discards that are avoided with a geometry shader. In the middle case, the number of discards remain the same.
On hardware that supports it, even a mesh shader pipeline may be preferable to generate quads to avoid the performance penalty of enabling geometry shaders. However, we did not consider this aspect in our work.

% Depth sorting
Gaussians with opacity require alpha blending and must be sorted by distance to the camera. The rendering order is determined by constructing and globally sorting an index buffer with one index per Gaussian. Compute shaders handle this process: one calculates distances from the camera, another uses a radix sort to reorder indices, and finally, indices are sent to the vertex shader to access attributes from a storage buffer.
%Since Gaussians have an opacity property, they must be rendered using alpha blending, which requires them to be sorted by distance to the camera. We can select which Gaussians are rendered and their order by constructing an index buffer with one index per Gaussian. We globally sort these indices, which determine the order in which Gaussians are drawn based on their distances from the camera. Three \gls{glsl} compute shaders are responsible for this. The distance between Gaussians and the camera is calculated in the first step. Then, a compute shader implementation of radix sort modifies the order of indices based on the calculated distances. The indices are provided to the vertex shader, allowing us to access the respective attributes from a storage buffer.

% Shaders
The math for drawing each Gaussian remains substantially the same from Kerbl \etal~\cite{3dgs}. The majority of the required calculations, described in Section~\ref{chap:vmem}, are performed once per Gaussian. They can, therefore, be performed in the vertex shader. The fragment shader is then responsible for shading the projected Gaussian within a quad based on a 2D covariance matrix.

\subsection{Limitations}
Our rendering concept has deficiencies that need addressing to scale to even larger scenes or enhance performance. These issues may overlap with those discussed in Section~\ref{chap:vmem} and will be detailed later. However, the most conceptual issues are discussed in the following.
%There are some deficiencies to our implementation, which may need to be cured to scale our solution to larger scenes or generally improve performance. The issues specific to implementation are separate but may overlap with issues in the method outlined in Section~\ref{chap:vmem}. We discuss these in more detail later.

% Page ID reductions
After rendering the proxy mesh image, it is reduced to a buffer indicating each page's visibility. Hable~\cite{hable} discusses the performance impact of \texttt{gl\_PrimitiveID} and suggests using the leading vertex to determine triangle primitive IDs. A compute shader reduces the visibility buffer to a list of IDs. To communicate additional \gls{lod} levels, the distance to the closest pixel is stored. Since finding this minimum can lead to race conditions, atomic operations are employed, impacting performance with concurrent thread updates. To enhance performance, it's better to minimize atomic operations, especially in global memory, and use faster shared memory within workgroups before accessing global memory.
%After rendering the proxy mesh with page IDs to an image, the image is reduced to a buffer, where each entry reflects whether the page in question is visible. John Hable~\cite{hable} describes the performance penalty introduced by \lstinline|gl_PrimitiveID|. He shows how the leading vertex for each triangle primitive can determine its ID instead. To reduce the visibility buffer to a list of IDs, a compute shader writes to the buffer. Besides simple visibility we also communicate the appropriate \gls{lod} level by storing the distance of the closest pixel of that page. Since finding such a minimum is sensitive to race conditions, an atomic operation is used. These operations impact performance if multiple threads simultaneously attempt to update the same entry. For better performance, atomic operations should be avoided, particularly in global memory. An improvement may include performing more operations in faster shared memory for workgroups before accessing global memory. 
We mitigate these shortcomings by rendering the visibility buffer at low resolution, which can cause distant pages to be missed. Mayer~\cite{vt_thesis} demonstrates how smaller visibility buffers improve performance with virtual texturing. However, unlike virtual texturing, overlapping Gaussians cause nearby pages to remain memory-bound.
%In practice, we currently overcome the impact of these shortcomings by rendering the visibility buffer at a low resolution. This can lead to missing pages, especially at greater distances. Mayer~\cite{vt_thesis} shows how reducing visibility buffer size can boost performance with virtual texturing. Unlike virtual texturing, nearby pages are likely to be linked due to overlapping Gaussians, which causes them to be required in memory regardless.

% Page table management (scale with pages, defer LOD switch, prioritize)
Our page table is a single array of entries, each managing a physical page, which may include multiple pages depending on the \gls{lod} level. We avoid hierarchical page tables found in operating systems to reduce indirections and potential page misses. Further work is needed to shrink page entries and test scalability with buffer size for rendering.
%Our page table consists of a single array of page table entries. A page table entry can store all the information required to manage a physical page, which may contain multiple pages, depending on the \gls{lod} level. We avoid a structure like a hierarchical page table, as used in operating systems, to reduce the number of indirections, and therefore potential page misses. More work can be done to reduce the size of page entries. Additional experiments are necessary to test this page table and ensure it scales well with the size of the buffer used for rendering. 
The page table entries, among other parts of the system, would also require modification to allow blending between \gls{lod} levels. During the transition, both levels are needed in memory simultaneously. 

Both \gls{lod} transitions and adding linked pages are less urgent than rendering visible Gaussian pages. The copy budget is limited by the staging buffer size and the time required. To maintain quality under high load, we prioritize important copies and delay \gls{lod} transitions to later frames.
%Both \gls{lod} transitions and the addition of linked pages are inherently less pressing than rendering the pages of Gaussians that are definitely visible based on their IDs being contained in the visibility buffer. The budget for copies is limited, artificially imposed by the size of the staging buffer but more generally caused by the time they take. For better quality under high load, more important copies should therefore be prioritized, deferring \gls{lod} transitions to later frames.

% No async streaming
Strict synchronization during frame rendering causes pipeline stalls. The \gls{gpu} could do more work but waits for task completion. Efforts to separate steps and perform tasks asynchronously could speed up rendering, especially data copying from \gls{cpu} to \gls{gpu}.
%As outlined in Section~\ref{sec:implementation_synchronization}, we introduce some strict synchronization between the major steps in rendering every frame. This introduces pipeline stalls, where the \gls{gpu} has capacity to perform more work but is unable to start on a new task until the last task is completed. With additional effort it may be possible to separate more of these steps, reducing the time taken to render a frame by completing some of the work asynchronously. Copying data from \gls{cpu} to \gls{gpu} is of special interest for such an optimization.

% Ignoring unified memory
Mobile devices, including those by Apple, generally contain integrated \gls{gpu}s nowadays. Memory is shared between \gls{cpu} and \gls{gpu}. To take advantage of this properly, modifying our concept could avoid streaming any data overall. Instead, memory could be mapped  from the file directly to the buffer used by the \gls{gpu} to render Gaussians.
%As described above, mobile devices, including those by Apple, generally contain integrated \gls{gpu}s. Memory is shared between \gls{cpu} and \gls{gpu}. To take advantage of this properly, our implementation could avoid streaming any data. Instead, memory could be mapped  from the file directly to the buffer used by the \gls{gpu} to render Gaussians.
% Software render
Finally, Gaussians are rendered using the \gls{gpu} hardware rasterizer. Kerbl \etal~\cite{3dgs} uses a software renderer in \gls{cuda}, breaking cross-platform portability in turn. Our concept neglects \gls{cuda} and is less restrictive. Nevertheless, the leverage of compute shaders for the proposed concepts could be advantageous on one or the other hardware.
%The final render of Gaussians is performed using an implementation that relies on the \gls{gpu}'s hardware rasterizer. In contrast, Kerbl \etal~\cite{3dgs} render Gaussians in a software renderer written in \gls{cuda}. While \gls{cuda} is not an option for us due to its proprietary nature, a very similar renderer could be created using compute shaders. This may lead to superior performance on at least some of the relevant hardware.

\begin{figure}[t]
	\centering
	\includegraphics[width=\linewidth,trim=40 40 0 40, clip]{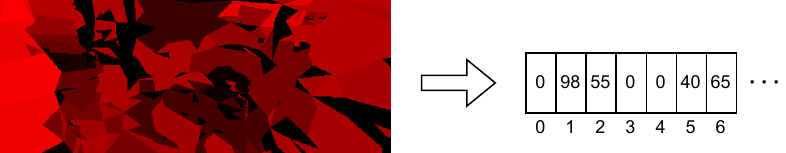}
    %\precapspace
	\caption{Rendered primitives are shaded based on the ID of their page. The buffer inside the compute shader holds one entry per page. The entry indicates that a page is visible by setting its respective entry to a value other than zero. Values encode depth for \gls{lod} level selection.}
    %\postcapspace
	\label{fig:implementation_reduction}
    \label{fig:vbuffer}
    \postcapspace
\end{figure}

\iffalse
\begin{figure*}[t]
    \centering
    \includegraphics[width=0.9\linewidth,trim=0 2 0 0,clip]{04/synchronization}
    \caption{High-level overview of the stages of rendering with virtual memory. Red bars indicate synchronization between \gls{cpu} and \gls{gpu}. Each step additionally depends on the previous step in the same lane. The tasks are not scaled with the time it takes to complete them.}
    \label{fig:synchronization}
\end{figure*}
\fi

\begin{figure*}[t]
    \centering
    \includegraphics[width=1\linewidth,trim=0 385 0 0,clip]{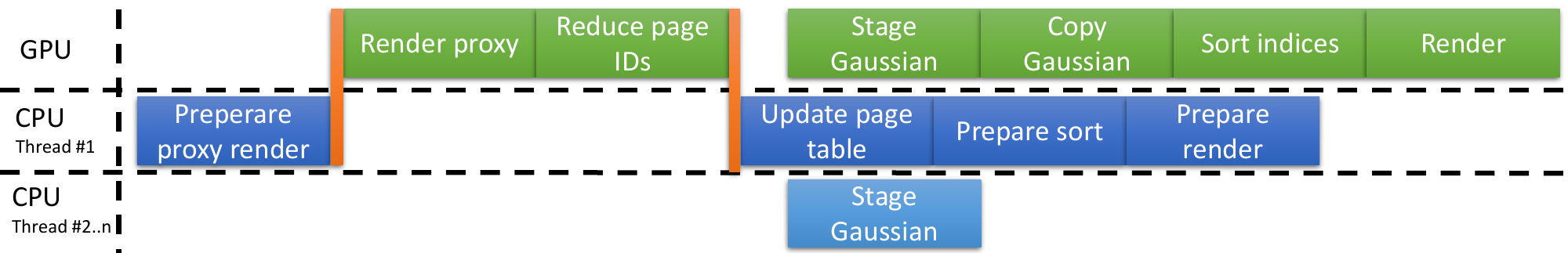}
    \caption{High-level overview of the stages of rendering with virtual memory. Orange bars indicate synchronization between \gls{cpu} and \gls{gpu}. Each step additionally depends on the previous step in the same lane. The tasks are not scaled with the time it takes to complete them.}
    \label{fig:synchronization}
    \postcapspace
\end{figure*}
% !TEX root = main.tex
\section{Evaluation}
\label{chap:evaluation}
% Intro; Explain what we care about (visuals [holes, popping, ...], performance, memory use [especially compared to full scene])
This section evaluates the method and implementation described in this work. In particular, we explore our results regarding visuals, memory use, and performance. 

\subsection{Details on Prototypical Implementation}
To better understand the visualization, we initially present important insight to our Vulkan-based implementation. The interested reader is referred to Haberl~\cite{haberl_msc2024} for a more in-depth description.

\subsubsection{Page determination details}
% Proxy mesh rendering (image type, ...)
We render the proxy mesh to a visibility buffer using a single-channel 32-bit unsigned integer image. This image can be much smaller than the final rendering while still delivering effective results. Page linking enhances this by connecting nearby pages, minimizing the effects of a smaller rendered image.
%To determine visible pages, we render the proxy mesh to a visibility buffer. We render an image with a single channel of 32-bit unsigned integers. This image may be significantly smaller than the final render while achieving good results. When using page linking, this is especially true since nearby pages are likely linked, reducing the impact of not drawing a smaller page. 

The vertex shader receives vertex positions and applies model and camera transformations used later in the scene. The fragment shader gets a buffer of page IDs for each face, with the input variable supplying the face index to output the corresponding page ID. Background and unassigned faces result in a zero page, ignored in later steps.
%The vertex shader is supplied with vertex positions and applies the same model and camera transformations, which will later be used for the actual scene. The fragment shader is additionally provided with a buffer of page IDs for each face. The \gls{glsl} input variable \lstinline|gl_PrimitiveID| supplies the face index, allowing the fragment shader to look up and output the corresponding page ID. The background and unassigned faces result in a zero page, which is ignored in subsequent steps.

% Compute shader for page ID reduction
After rendering the proxy mesh, we reduce the data by creating a buffer indicating required pages. A compute shader updates each pixel's list entry in parallel, encoding depth as integers that decrease with greater depth. Using the \texttt{atomicMax()} function, only the closest depth value is written to the buffer. We iterate through linked pages per pixel to set buffer entries. Once transferred to CPU memory, the array shows zeroes for unneeded and depth values for needed pages. Selection is based on depth value \gls{lod}, as visualized in Figure~\ref{fig:implementation_reduction}. Transitions between detail levels are adjusted at runtime based on used to free memory ratio, aiming for 50-80\% usage. Thresholds are moved if outside this range, increasing step size by 1\% after successive similar adjustments or decreasing it by 1\% if direction changes quickly.
%Once the proxy mesh has been rendered, we reduce the contained data. We create a buffer where each entry signals whether a page is required. Using a compute shader we can update the list entry for each pixel in parallel. We write the encoded depth to the pixel to allow \gls{lod} selection. The depth is encoded to represent it in integers, where the values decrease with increasing depth. Using the \lstinline|atomicMax()| function, we write only the largest value, and therefore the closest depth, to the buffer. We iterate over all linked pages for each pixel and set the required buffer entries using the same depth. Once copied to CPU memory, the resulting array contains zeroes for pages that are not required and depth values for required ones. The desired \gls{lod} can be chosen based on the depth value. This process, from rendering to reduction, is visualized in Figure~\ref{fig:implementation_reduction}.

%The thresholds for these transitions between levels of detail are updated at runtime and adjusted to the scene contents, scale, and camera view. This is achieved by monitoring the ratio of used to free memory. We target a ratio of between 50\% and 80\% memory usage, outside of which we move the thresholds. If multiple threshold adjustments in the same direction happen in quick succession, we increase the step size by 1\%. However, if a move in one direction is soon followed by a move in the other, we decrease the step size by 1\%.

\subsubsection{Transfer}
The buffer of Gaussians is memory-mapped on the \gls{cpu}, enabling buffer access while file reads are managed by the OS. Discrete \gls{gpu}s' large memory segments are not directly accessible from the \gls{cpu}. We copy these to host-visible memory for writing, despite its small size and slower \gls{gpu} access. This staging buffer is vital for discrete \gls{gpu}s, as they have independent memory separate from the \gls{cpu}, connected typically via \gls{pcie}. Integrated \gls{gpu}s, used in mobile devices, share memory with the \gls{cpu}, making all accessible memory host-visible but with lower throughput. The implementation completes all memory transfers before further steps, with potential enhancements through separate GPU queues for reduced framerate impact.

%The entire buffer of Gaussians is memory-mapped on the \gls{cpu}. We can, therefore, access the buffer, leaving file reads to the operating system. In contrast, large sections of memory in discrete \gls{gpu}s cannot be accessed from the \gls{cpu}. We first copy it to a buffer in host-visible memory, which allows us to write to it. Host-visible memory is generally small and slower for the \gls{gpu} to access. Thus, we use this buffer for staging, immediately copying the data to a larger buffer in a heap that only the \gls{gpu} can read from.

%Note that this use of a staging buffer is optimized and, in fact, necessary only for discrete \gls{gpu}s. These are dedicated components with their own memory, entirely separate from the \gls{cpu}, connected only via an interconnect (typically \gls{pcie}). Integrated \gls{gpu}s, such as those used in mobile devices, share their memory with the \gls{cpu}. As a result, all memory that the \gls{gpu} can access is host-visible, though throughput is much lower.

%Our implementation requires completing all memory transfers before proceeding with later steps. Further improvements can be made by transferring data in a separate queue on the GPU, which would further reduce the impact on framerates.

\subsubsection{Synchronsitaion}
Work is largely performed on the \gls{gpu}. Steps in Figure~\ref{fig:synchronization} benefit from parallel execution but require synchronization. The page update begins with recording commands to draw the proxy mesh and reduce page IDs, separated by a pipeline barrier to ensure memory is flushed before processing results. These commands are submitted to a graphics and compute-capable queue. We use a fence to wait before updating the page table on the \gls{cpu}. Data transfer to the staging buffer starts during the page table update, with multi-threaded transfers handled by the drivers. After updating, Gaussians are copied to the second command buffer, also containing depth sorting and drawing, separated by a barrier. Commands are recorded during data transfer in separate threads. Depth sorting via compute shaders requires nine dispatches with metadata in a uniform buffer, defined by a push constant. Lean barriers prevent pipeline stalls. Post-dispatch, depth results move to the index buffer. Rendering the final \gls{3dgs} scene uses the updated index buffer. After updating the page table and completing transfers, command buffer submission triggers a semaphore, presenting the image. Two frames can be processed simultaneously, with work on the \gls{cpu} while the \gls{gpu} presents the previous frame.

\subsubsection{Adaptations for Mobiles}
\begin{figure}[t]
	\centering
	\includegraphics[width=1\linewidth,trim=15 245 80 0, clip]{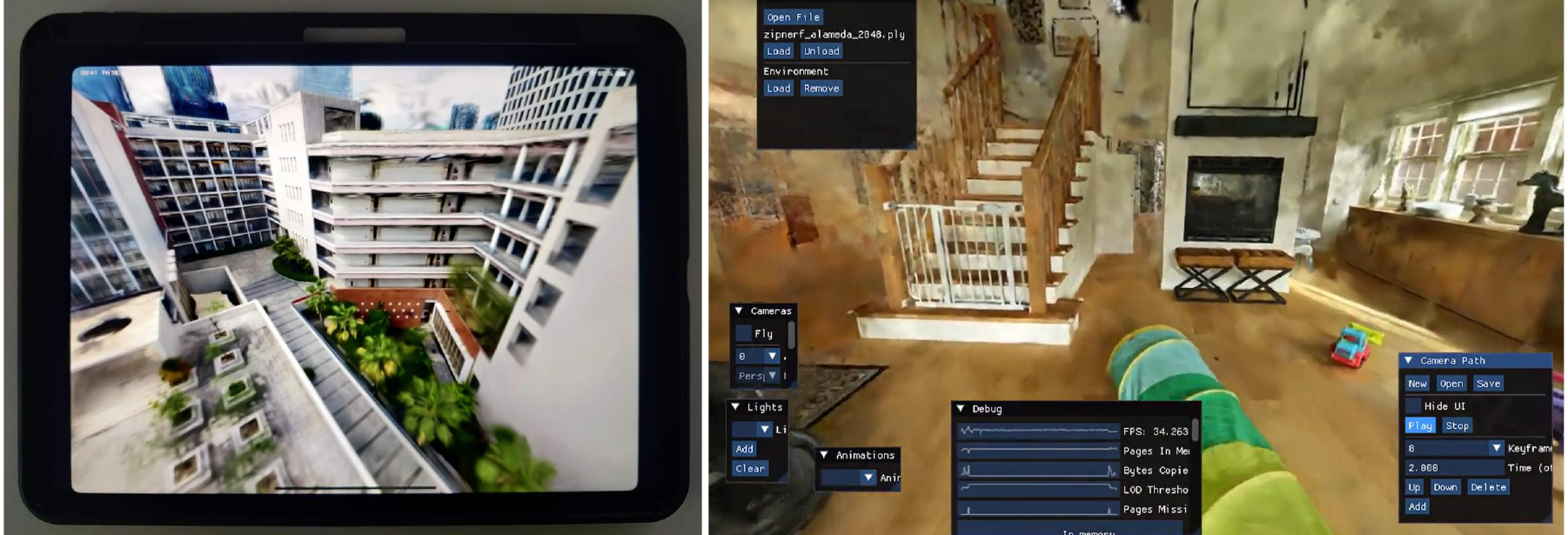}
	\caption{Our implementation running on an iPad Pro. The application is not specifically optimized for mobile devices with integrated \gls{gpu}s. It uses the same Vulkan code for rendering with MoltenVK as a translation layer to the Metal API.}
	\label{fig:mobile_device}
    \postcapspace
\end{figure}

We port our real-time renderer to iOS to test its viability on mobile devices. Since Apple prefers the Metal API over Vulkan, we use MoltenVK from the Khronos Vulkan Portability Initiative~\cite{vulkan_portability} to bridge Vulkan and Metal, allowing our existing Vulkan code to run on macOS and iOS. Despite some extension support differences with Windows, the essential extensions required are supported. The main differences between desktop OS (Windows, macOS) and iOS relate to file handling. An iOS app bundle includes an executable and resources, while it can read from and write to an app-specific, sandboxed directory. We package all shaders and test scenes into the app bundle, using the POSIX function \texttt{mmap(...)} for memory mapping. Apple devices fully support POSIX, and system interactions are feasible via the C++ standard library. While we adjust Vulkan and file-related code to ensure our implementation compiles and runs, we do not optimize for mobile devices or iOS specifically. Figure~\ref{fig:mobile_device} shows the application on a 3rd Generation iPad Pro.

%To evaluate our method on a mobile device we port our real-time renderer to iOS. Our aim is to ensure the techniques used in our implementation (e.g., memory-mapped files) are viable on mobile devices.

%Apple devices do not support Vulkan in favor of Apple's own Metal API. MoltenVK, a part of the Khronos Vulkan Portability Initiative~\cite{vulkan_portability}, provides a layer between Vulkan and Metal. It enables us to use our existing Vulkan code on macOS and iOS. There are some differences in the extension support to common Vulkan drivers on Windows but all extensions that are important to our application are supported. 

%The biggest differences between desktop operating systems (Windows, macOS) and iOS can be found when working with files. An iOS app bundle contains an executable and resources to read from. The app is also able to read from and write to an app-specific, sandboxed, directory. We choose to include all required shaders and our test scenes in the app bundle. To open scenes using memory mapping we rely on the POSIX function \texttt{mmap(...)}. Apple devices, including mobile, are fully capable POSIX systems. Interactions with the operating system are possible through the C++ standard library.

%We make these changes to Vulkan and file related code that are fundamentally necessary to compile and run our implementation. No additional effort is made to optimize for mobile devices with integrated \gls{gpu}s in general or iOS in particular. Figure~\ref{fig:mobile_device} shows the application running on a 3rd Generation iPad Pro.

\begin{figure*}[t]
    \centering
    \includegraphics[width=1\linewidth,trim=0 412 0 0, clip]{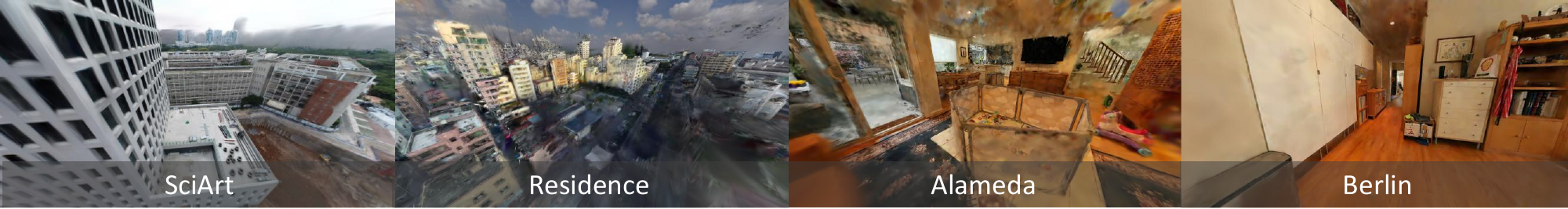}
    \caption{Example images of the evaluated scenes: \textit{SciArt} and \textit{Residence} are taken from the UrbanScene3D~\cite{urbanscene3d} dataset and trained with VastGaussian~\cite{vast_gaussian}. And \textit{Alameda} and \textit{Berlin} are from Zip-NeRF~\cite{zipnerf} and trained with Nerfstudio~\cite{nerfstudio}.}
    \label{fig:scenes}
    \postcapspace
\end{figure*}

\subsubsection{Build Details and Test Hardware}
% Describe implementation
As mentioned, our renderer is implemented in C++ using Vulkan, which uses the hardware rasterizer. The evaluation is done on Windows computer and we compile our application with \gls{msvc}. The build type is set to \texttt{RelWithDebInfo} to optimize the binary for best performance. We disable any Vulkan validation layers during our evaluation. 

% Describe hardware
Our testing hardware includes an NVIDIA GeForce GTX 1070 \gls{gpu} and Intel Core i7-4770k \gls{cpu}, using \gls{ddr} system memory to transfer Gaussian via \gls{pcie} 3.0. Though not state of the art, this setup is expected to remain common among consumers~\cite{tatarchuk}. Thus, good performance on such devices is crucial for \gls{3dgs}'s adoption.
%The hardware we run our tests on consists of an NVIDIA GeForce GTX 1070 \gls{gpu} and an Intel Core i7-4770k \gls{cpu}. We use \gls{ddr} system memory and transfer Gaussian via \gls{pcie} 3.0. Such a system is far from the current state of the art, and better performance can be expected in all aspects of modern systems. Tatarchuk~\cite{tatarchuk} expects hardware with performance similar to our test system to remain widespread for the foreseeable future among consumers. Therefore, good performance on these devices is imperative for the adoption of \gls{3dgs} in end-user applications.
% Mobile
We also test our method on an Apple iPad Pro, model A2377 with an Apple M1 \gls{soc}, featuring \gls{cpu} and integrated \gls{gpu}, to assess performance impacts on mobile devices and guide future development regarding integrated \gls{gpu}s.
%To explore the performance implications of our method on mobile devices, we additionally perform some tests on an Apple iPad Pro. Specifically we choose model number A2377 with an Apple M1 \gls{soc} combining \gls{cpu} and integrated \gls{gpu}. We aim to determine the performance implications of integrated \gls{gpu}s for our implementation and guide future work.

\subsection{Qualitative Analysis} %Former Scene Eval
We evaluate various \gls{3dgs} scenes of different sizes, defining a camera path for each using a perspective camera with a 90° \gls{fov}. The camera moves at a fixed speed, interpolating its position and orientation between checkpoints. Paths are designed to encompass diverse scenarios, including close-ups and distant overviews, thoroughly navigating the scene to engage multiple Gaussian pages at varying detail levels, thereby showcasing the method's strengths and weaknesses.
%We evaluate a variety of \gls{3dgs} scenes of different sizes. For each, we define a camera path. We create a perspective camera with a 90° \gls{fov}. The camera linearly interpolates its translation and rotation between defined checkpoints at a fixed speed. Each path aims to create a variety of different situations. These include both close-ups and overviews at large distances. The scene is navigated thoroughly to require many different pages of Gaussians at differing levels of detail. This provides a thorough overview of the strengths and weaknesses of both the method and implementation.

As the camera follows its path, we track key statistics per frame: memory used by active Gaussians, and the time to render the proxy mesh, list pages, update the page table, copy data, sort, and render Gaussians.
%While the camera is following its predefined path, we measure relevant statistics. In each frame, we determine the amount of memory filled with actively required Gaussians. We measure the time it takes to render the proxy mesh and reduce it to a list of pages, update the page table, copy data, sort Gaussians, and render them.

\begin{table}[t]
\footnotesize
    \centering
    \footnotesize
	\begin{tabular}{|p{1cm}||p{1.2cm}|p{0.8cm}|p{0.8cm}|p{1cm}|p{1.2cm}|}
		\hline
		Scene & Initial Size [MiB] & Pages & Links & File Size [MiB] & Pre processing [s] \\
		\hline\hline
		SciArt & 804.9 & 3778 & 49576 & 1505.5 & 2600 \\
		\hline
		Residence & 2214.6 & 4954 & 65856 & 4281.5 & 4579 \\
		\hline
		Alameda & 353.9 & 730 & 10664 & 630.9 & 1169 \\
		\hline
		Berlin & 236.5 & 492 & 6444 & 425.2 & 921 \\
		\hline
	\end{tabular}
	\caption{
    Evaluation statistics for scenes with a page size of 2048 include initial sizes post-reconstruction, followed by updates such as padding and added \gls{lod} levels, affecting final sizes. Preprocessing times for each scene are also provided.
    %Statistics for the scenes used for evaluation, with a page size of 2048. The initial size is measured after reconstructing the scenes, before any modifications are made. Then, padding is added and additional \gls{lod} levels are generated. These modifications impact the final size of the scene. The preprocessing time for each scene is also listed.
    }
	\label{tab:scene_data}
    %\postcapspace
\end{table}

%\subsection{Scenes}

\iffalse
\begin{figure}[t]
    \centering
    \begin{subfigure}[t]{.49\linewidth}
        \includegraphics[height=2.2cm]{05/scenes/scene_sciart}
        \caption{SciArt}
    \end{subfigure}
    \begin{subfigure}[t]{.49\linewidth}
        \includegraphics[height=2.2cm]{05/scenes/scene_residence}
        \caption{Residence}
    \end{subfigure}
    \begin{subfigure}[t]{.49\linewidth}
        \includegraphics[height=2.2cm]{05/scenes/scene_alameda}
        \caption{Alameda}
    \end{subfigure}
    \begin{subfigure}[t]{.49\linewidth}
        \includegraphics[height=2.2cm]{05/scenes/scene_berlin}
        \caption{Berlin}
    \end{subfigure}
    \vspace{-10pt}
    \caption{Rendered image for each of the four scenes used for evaluation.}
    \label{fig:scenes}
\end{figure}
\fi

% Introduce datasets
Large scenes are expected to make extensive use of \gls{lod}, while complex scenes with potential for occlusions allow us to eliminate many Gaussians from rendering.
We therefore choose a selection of large-scale scenes trained with the methods introduced by Lin \etal~\cite{vast_gaussian}. 
%C-out They can create large scenes with great detail. However, especially after \gls{lod} generation, we observe artifacts. These include large Gaussians, which do not adhere well to surfaces. 
Kerbl \etal~\cite{hierarchical_3dgs} present a similar dataset in their work on large-scale rendering, but it is not available at the time of writing. Smaller scenes with obstructions are created from images by Barron \etal~\cite{zipnerf} and trained using Nerfstudio~\cite{nerfstudio}. 
%C-out These scenes contain many artifacts, including floaters and holes. However, it provides insight into how virtual memory can be used with indoor scenes, which feature many occlusions. 
The selected scenes are listed in Table~\ref{tab:scene_data}, with an image of each in Figure~\ref{fig:scenes}.
Datasets in novel view synthesis often consist of images of a single central object captured from all angles. They contain negligible amounts of occlusion and, therefore, generally do not benefit significantly from virtual memory. We omit testing for these types of scenes.

Table~\ref{tab:scene_data} presents statistics per scene with a 2048 page size. Page numbers correlate with Gaussian counts despite padding. Adaptive \gls{lod} allows the same buffer for scenes with varying pages, aiming to maintain stable frame times and memory. When pages exceed the buffer, higher \gls{lod} levels decrease quality. Alternatively, positioning the camera to capture fewer pages permits rendering at lower \gls{lod} levels, improving results.
%Table~\ref{tab:scene_data} lists various statistics for each scene after preprocessing with a page size of 2048. The number of pages in a scene is closely related (though not perfectly due to padding) to the number of Gaussians. When using adaptive \gls{lod}, the same size buffer may be used for different scenes with different page counts. We therefore aim to keep frame times and memory utilization from being affected. Higher \gls{lod} levels are used when a large number of pages don't fit into the buffer, reducing visual quality. Alternatively, the camera may be placed such that only a small subset of pages are visible, which can be rendered at lower \gls{lod} levels. Clearly, this is preferable to achieve good results.
Creating a \gls{3dgs} scene involves steps like \gls{sfm}, training, and preprocessing, which are not deterministic and may lead to different statistics on recreation. We consistently find an average of 13-15 links per page across scenes. Similar to mipmaps, halving the page size with each \gls{lod} level nearly doubles the final size.
%The steps involved in creating a \gls{3dgs} scene, such as \gls{sfm}, training, and preprocessing, are generally not deterministic. The presented statistics may, therefore, change when recreating a scene. We observe a consistent average number of links per page of around 13-15 for all scenes. Like with mipmaps, as we half the page size with each level when generating \gls{lod}s, the final size approaches double the initial size. 
The overhead we introduce is negligible compared to the size of the original scene. We store the proxy mesh with its page IDs, metadata, and page links in a single file. This file stays below 20 MiB in all listed scenes.

Preprocessing times excluding slice rendering are given in the last column in Table~\ref{tab:scene_data}. Factors like size, scene structure, and hardware impact these times. The largest scene, residence, requires over an hour to preprocess, with 85\% spent on \gls{lod} generation. Its structure makes k-means converge slowly. For most scenes, page assignment, linking, and \gls{lod} generation each make up half the preprocessing time. Excluding residence, each page adds about half a second. \gls{lod} generation, being highly parallel, can be accelerated by reducing iterations or using a faster \gls{cpu} with more cores.
%Preprocessing times for each scene are also shown in Table~\ref{tab:scene_data}. This excludes slice rendering which takes just seconds. Multiple factors greatly influence these timings, including the size, scene structure, and hardware. The largest scene, residence, takes slightly longer than an hour to preprocess. The vast majority of this time, almost 85\%, is spent generating \gls{lod}s. The structure of this scene makes k-means converge particularly slowly. For most scenes page assignment and linking, and \gls{lod} generation each make up about half of the overall preprocessing time. When disregarding residence, each page adds about half a second overall. \gls{lod} generation is highly parallel and can be sped up by reducing the maximum number of iterations or running it on a faster \gls{cpu} with more cores.

\begin{figure}[t]
    \centering
    \includegraphics[width=\linewidth,trim=0 288 380 50, clip]{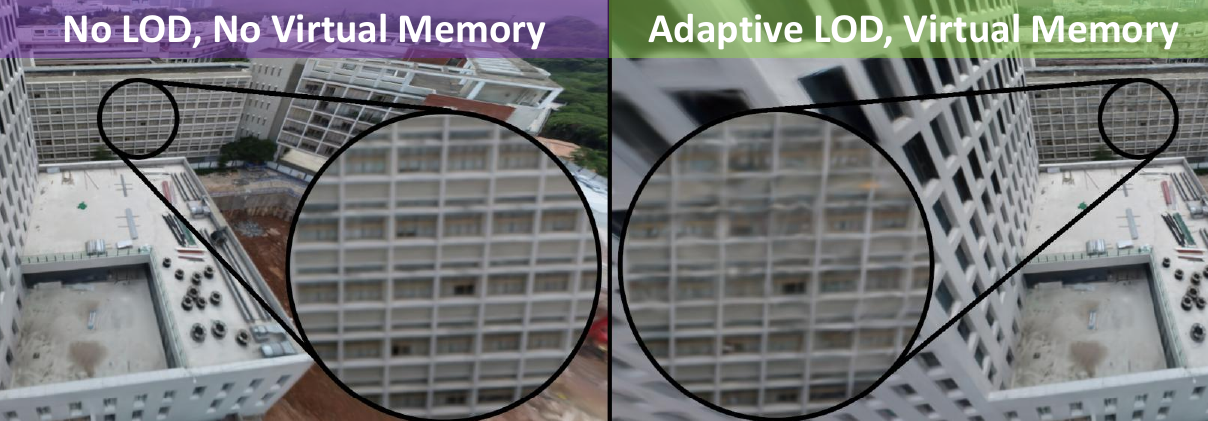}
    \caption{Similar quality, smaller memory footprint: To avoid creating holes due to lack of available memory, the scene is rendered entirely without virtual memory. Virtual memory with adaptive \gls{lod} is enabled. Close to the camera, the original Gaussians are rendered. %As the distance from the camera increases the right image renders higher levels with fewer Gaussians per page.
    }
    \label{fig:demo_lod}
    \postcapspace
\end{figure}

%\subsection{Demonstration}
% Basic introduction to results
\iffalse
\begin{figure}[t]
    \centering
    \includegraphics[width=0.49\linewidth]{05/demo/demo_frustum}
    \includegraphics[width=0.49\linewidth]{05/demo/vis_comp_full}
    \caption{To render the image on the right, only a subset of all Gaussians are rendered. The left image shows the same scene from a different perspective. Gaussians outside the view frustum and those hidden from the camera by occlusions are not rendered.}
    \label{fig:demo_frustum}
\end{figure}
\fi

\begin{figure*}[t]
    \centering
    \includegraphics[width=1\linewidth,trim=0 420 50 0, clip]{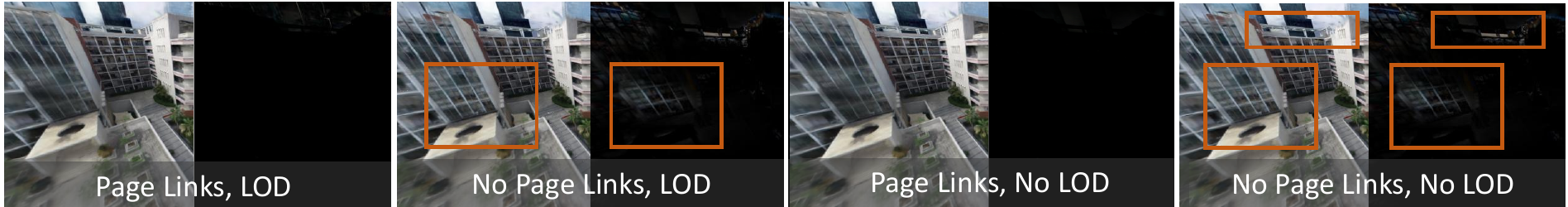} \\
    (a) \hspace{4cm} (b) \hspace{4cm} (c) \hspace{4cm} (d)
    \vspace{-5pt}
    \caption{Images with the same camera position and settings, rendered with ablations of our method (left). Images on the right compare these to the same scene rendered virtual memory. Black indicates no difference, pixel brightness indicates how large the difference is.}
    \label{fig:vis_comp_diff}
    \label{fig:vis_comp_diff_d}
    \label{fig:vis_comp_diff_c}
    \label{fig:vis_comp_diff_b}
    \label{fig:vis_comp_diff_a}
    \postcapspace
\end{figure*}

Page assignment is solely position-based. Gaussians at the building's top, likely facing upward, should be rendered even if they're not in the visibility buffer. With page links, pages at the front and top are linked due to overlap. Figure~\ref{fig:demo_lod} contrasts scene rendering without virtual memory against using it with \gls{lod}. Gaussians further from the camera use higher \gls{lod} levels. During preprocessing, Gaussians within a page are combined to halve those of the lower level. At runtime, memory use determines the distance from the camera where \gls{lod} level changes. Higher levels are applied further away, slightly lowering visual quality, but clustering such pages in a render buffer boosts performance.

The preprocessed scene is loaded to prepare for real-time rendering. The scene is rendered using virtual memory with the implementation described in this work. To determine their impact, we enable the methods presented separately (linking, \gls{lod}, etc.).
In Table~\ref{tab:psnr_ssim} we calculate \gls{psnr} and \gls{ssim} between an image rendered with and without virtual memory. We render the same image with various ablations of our method. Figure~\ref{fig:vis_comp_diff} contains all image variations and their differences from the scene rendered without virtual memory. None of the configurations are memory-constrained; all required pages fit in memory.

When \gls{lod} is disabled but page links are used, almost all of the 500 available pages are used. With buffer usage above 80\%, adaptive \gls{lod} reduces the distances for level transitions. Image quality suffers as a result, with a sharp drop in \gls{psnr}. With our test setup the ranges adaptive \gls{lod} lands on to satisfy conditions on memory use are not deterministic. When page links are not used, noticable artifacts appear in the image which is reflected in the image comparison scores. These scores listed in Table~\ref{tab:psnr_ssim}, as well as the right image in Figure~\ref{fig:vis_comp_diff_c}, indicate the best quality with page links enabled and \gls{lod} disabled. This makes sense given that page links increase the number of required pages to produce better results while \gls{lod} diminishes quality to regain memory. However, this quality can only be achieved because all required pages, even with the additional pages requested due to page links, can fit into the 500 page buffer. If this were not the case, additional artifacts would be introduced. Disabling page linking (Figure~\ref{fig:vis_comp_diff_b} (b) and (d)) leads to the expected artifacts explored previously. Enabling both methods can be thought of as balancing quality and memory usage. In Figure~\ref{fig:vis_comp_diff_a}, there are little to no differences for parts of the scene close to the camera. Page links fulfill their part of contributing pages not visible on the proxy mesh but overlapping with pages that are. Simultaneously, when those pages are distant, a different \gls{lod} level is used, which reduces quality slightly.

\begin{table}[t]
\footnotesize
	\centering
	\begin{tabular}{|l||c|c|}
		\hline
		ArtSci & \gls{psnr} [dB] & \gls{ssim}\\
		\hline\hline
		With page links, with \gls{lod} & 42.89 & 0.99 \\
		\hline
		Without page links, with \gls{lod} & 35.02 & 0.97 \\
		\hline
		With page links, without \gls{lod} & 47.19 & 0.99 \\
		\hline
		Without page links, without \gls{lod} & 35.02 & 0.97 \\
		\hline
	\end{tabular}
	\caption{Comparison between images rendered without virtual memory and with ablations of our virtual memory solution.}
	\label{tab:psnr_ssim}
        \postcapspace
\end{table}

\iffalse
\begin{figure}[t]
	\centering
	\begin{subfigure}[t]{\linewidth}
		\includegraphics[width=.49\linewidth]{05/demo/vis_comp_full}
		\includegraphics[width=.49\linewidth]{05/demo/vis_comp_full_diff}
		\caption{With page links, with \gls{lod}}
		\label{fig:vis_comp_diff_a}
	\end{subfigure}\\
	\begin{subfigure}[t]{\linewidth}
		\includegraphics[width=.49\linewidth]{05/demo/vis_comp_nolink}
		\includegraphics[width=.49\linewidth]{05/demo/vis_comp_nolink_diff}
		\caption{Without page links, with \gls{lod}}
		\label{fig:vis_comp_diff_b}
	\end{subfigure}\\
	\begin{subfigure}[t]{\linewidth}
		\includegraphics[width=.49\linewidth]{05/demo/vis_comp_nolod}
		\includegraphics[width=.49\linewidth]{05/demo/vis_comp_nolod_diff}
		\caption{With page links, without \gls{lod}}
		\label{fig:vis_comp_diff_c}
	\end{subfigure}\\
	\begin{subfigure}[t]{\linewidth}
		\includegraphics[width=.49\linewidth]{05/demo/vis_comp_nolinknolod}
		\includegraphics[width=.49\linewidth]{05/demo/vis_comp_nolinknolod_diff}
		\caption{Without page links, without \gls{lod}}
		\label{fig:vis_comp_diff_d}
	\end{subfigure}\\
    \precapspace
	\caption{Images with the same camera position and settings, rendered with ablations of our method (left). Images on the right compare these to the same scene rendered virtual memory. Black indicates no difference, pixel brightness indicates how large the difference is.}
    \postcapspace
	\label{fig:vis_comp_diff}
\end{figure}
\fi

\begin{figure}
    \centering
    \includegraphics[width=1\linewidth,trim=0 361 500 0, clip]{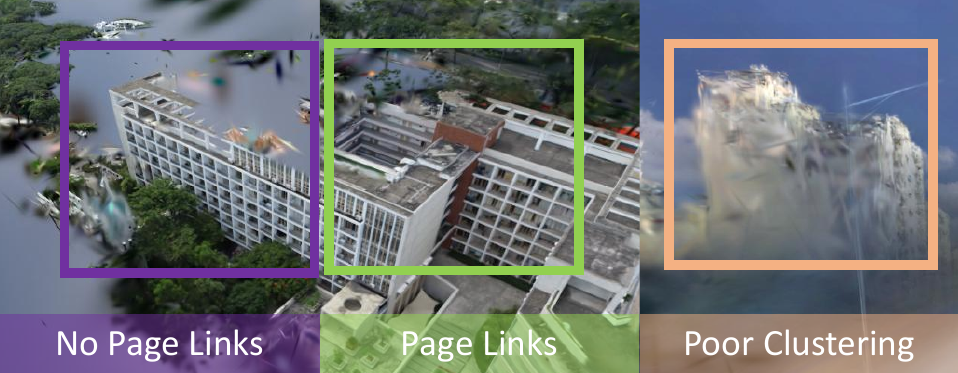}
    \caption{No Page Links leading only the pages of Gaussians determined to be visible based on the visibility buffer are transferred and rendered. Enabled page links, causing significantly more pages to be rendered based on page overlaps. Poor clustering of Gaussians to merge can create artifacts, which becomes obvious when high \gls{lod} levels are used close to the camera.}
    \label{fig:demo_link_problems}
    \label{fig:demo_lod_problems}
    %\postcapspace
\end{figure}

\paragraph{Limitations}
Figure~\ref{fig:demo_link_problems} demonstrates the issues our current page link method presents. No special care is taken to avoid reducing the number of page overlaps during preprocessing. In extreme cases, this can negate the occlusion culling effect achieved by the visiblity buffer entirely. While \gls{lod} might mitigate overhead, higher \gls{lod} levels may lower quality. \gls{lod} lowers Gaussian page quality, with artifacts appearing when merging over distances. Though minor when far from the camera, these artifacts become significant near the camera at high \gls{lod} levels, as shown in Figure~\ref{fig:demo_lod_problems}. Reducing page overlaps and adjusting Gaussian scales can decrease artifacts. Our method does not blend levels to hide popping during \gls{lod} level transitions, a feature left for future versions with \gls{3dgs}'s alpha blending capability.

\subsection{Quantitive Analysis}
%Several factors impact performance with our method, presenting various trade-offs. 

% Cache locality
Grouping Gaussians together enhances cache locality, increasing cache hits during rendering, especially with virtual memory. Although page order doesn't reflect Gaussian proximity, streaming visible pages to a smaller buffer brings pages closer.
%Creating pages of Gaussians causes those in close proximity to be stored near each other in the file and subsequently in memory. This improves cache locality by making cache hits more likely during rendering. When rendering with virtual memory, this effect is even more pronounced. Gaussians lie close to each other within each page, but the order of pages is not related to their relative positions. However, when streaming the visible subset of pages to the smaller buffer on the \gls{gpu}, they are likely to be closer to each other.
% Compare different page sizes
Larger page sizes raise position variance and cause more Gaussians to be falsely marked visible due to a less atomic visibility buffer, while smaller sizes add overhead to page table management and memory transfers.
%The chosen page size, defined by the maximum number of Gaussians grouped in each page, naturally changes the effect on cache locality. Larger page sizes increase the variance in positions. Additionally, more Gaussians are incorrectly marked as visible since the visibility buffer is less atomic. Small page sizes, however, increase the overhead for page table management and memory transfers. 
%Experiments confirm these assumptions. 
Table~\ref{tab:pagesize_impact} compares timings of selected rendering stages for different page sizes. Other stages are largely unaffected. Notably, the time gained updating the page table at larger page sizes does not seem to outweigh the penalty of unnecessary Gaussians and reduced cache locality, at least for large scenes. To reduce the impact of cache locality, future implementations may additionally choose to reorder Gaussians within each page based on the Morton order. Table~\ref{tab:pagesize_impact} indicates, that a 2048 page size is appropriate to balance cache locality with page table management overhead.
%Based in part on the data in Table~\ref{tab:pagesize_impact} we select a page size of 2048 for our experiments in this section. We find this is a good balance between cache locality and the overhead associated with page table management.
% Buffer size
The size of the staging and rendering buffers is crucial. A larger staging buffer permits longer delays, as copying Gaussians blocks depth-sorting and rendering. Conversely, if the buffer is too small for all required pages yet fails to present pages in \gls{gpu}, artifacts like holes and popping occur. Our staging buffer is set to a size of $\sim$18.4MiB, equal to 40 pages of 2048 Gaussians.
%Another variable is the size of the staging buffer and the buffer used for rendering. A large staging buffer allows for long delays since copying Gaussians blocks depth-sorting and rendering in our implementation. However, if the staging buffer is too small to copy all required pages but does not present pages in \gls{gpu}, noticeable holes and popping artifacts occur. For experiments, we choose a staging buffer size of $\sim$18.4MiB, equivalent to 40 pages of 2048 Gaussians.

\begin{table}[t]
    \centering
    \footnotesize
    \begin{tabular}{|c||c|c|c|}
        \hline
        Page Size & Page Table Update [ms] & Depth Sort [ms] & Render [ms]\\
        \hline\hline
        2048 & 0.9 & 2.0 & 8.0 \\
        \hline
        4096 & 0.5 & 3.0 & 9.7 \\
        \hline
        8192 & 0.3 & 6.7 & 10.6 \\
        \hline
    \end{tabular}
    \caption{Comparison of median time taken for selected rendering stages with different page sizes. In all tests, the camera follows the same predefined path. As the page size increases, page table updates shorten, while depth sorting and rendering take longer.}
    \label{tab:pagesize_impact}
    \postcapspace
\end{table}

\begin{table}[t]
	\centering
    \footnotesize
	\begin{tabular}{|c||p{1cm}|p{1cm}||p{1cm}|p{1cm}|}
		\hline
		\multirow{2}{*}{Buffer Size [Pages]} & \multicolumn{2}{c||}{Without \gls{lod}} & \multicolumn{2}{c|}{With \gls{lod}} \\
		\cline{2-5}
		& Render [ms] & Missing Pages & Render [ms] & Missing Pages \\
		\hline\hline
		250 & 3.8 & 549 & 5.0 & 0 \\
		\hline
		500 & 6.6 & 263 & 6.7 & 0 \\
		\hline
		1000 & 8.1 & 0 & 8.4 & 0 \\
		\hline
	\end{tabular}
	\caption{Comparison of median render times and number of missing pages for different buffer sizes with and without \gls{lod}. Buffer sizes are specified as a multiple of the page size. Both with and without \gls{lod}, the render time increases with the buffer size. However, \gls{lod} reduces the necessary memory to contain the required pages.}
	\label{tab:buffer_size_impact}
    %\postcapspace
\end{table}

%The rendering buffer size significantly affects performance. It must accommodate visible pages. To address performance issues from rendering many Gaussians, \gls{lod} is used. \textbf{Our implementation modifies the distance at which \gls{lod} levels change based on the amount of memory used.} \textbf{In other words, it aims to fit all required pages into the buffer to avoid holes.} Large buffers also allow rendering at low \gls{lod} levels, harming performance. 
The rendering buffer size significantly affects performance. The buffer must be large enough to hold visible pages. \gls{lod} addresses performance issues from rendering numerous Gaussians: \textbf{Our implementation adjusts \gls{lod} levels based on memory usage.} \textbf{In other words, it aims to fit all required pages into the buffer to avoid holes.} As a side effect, large buffers allow many Gaussians to be rendered at low \gls{lod} levels (more details), negatively impacting performance. Table~\ref{tab:buffer_size_impact} examines the link between buffer size, render times, and missing pages. Without \gls{lod}, smaller buffer sizes are too small to hold all visible pages. As a result, render times increase when increasing the buffer size. When using adaptive \gls{lod}, distances are chosen to avoid any holes due to missing pages. \textbf{As a result, a smaller buffer can render the same scene without artifacts.}

\subsubsection{Memory Usage}

\begin{figure}[t]
    \centering
    \includegraphics[width=1\linewidth,trim=5 112 250 30,clip]{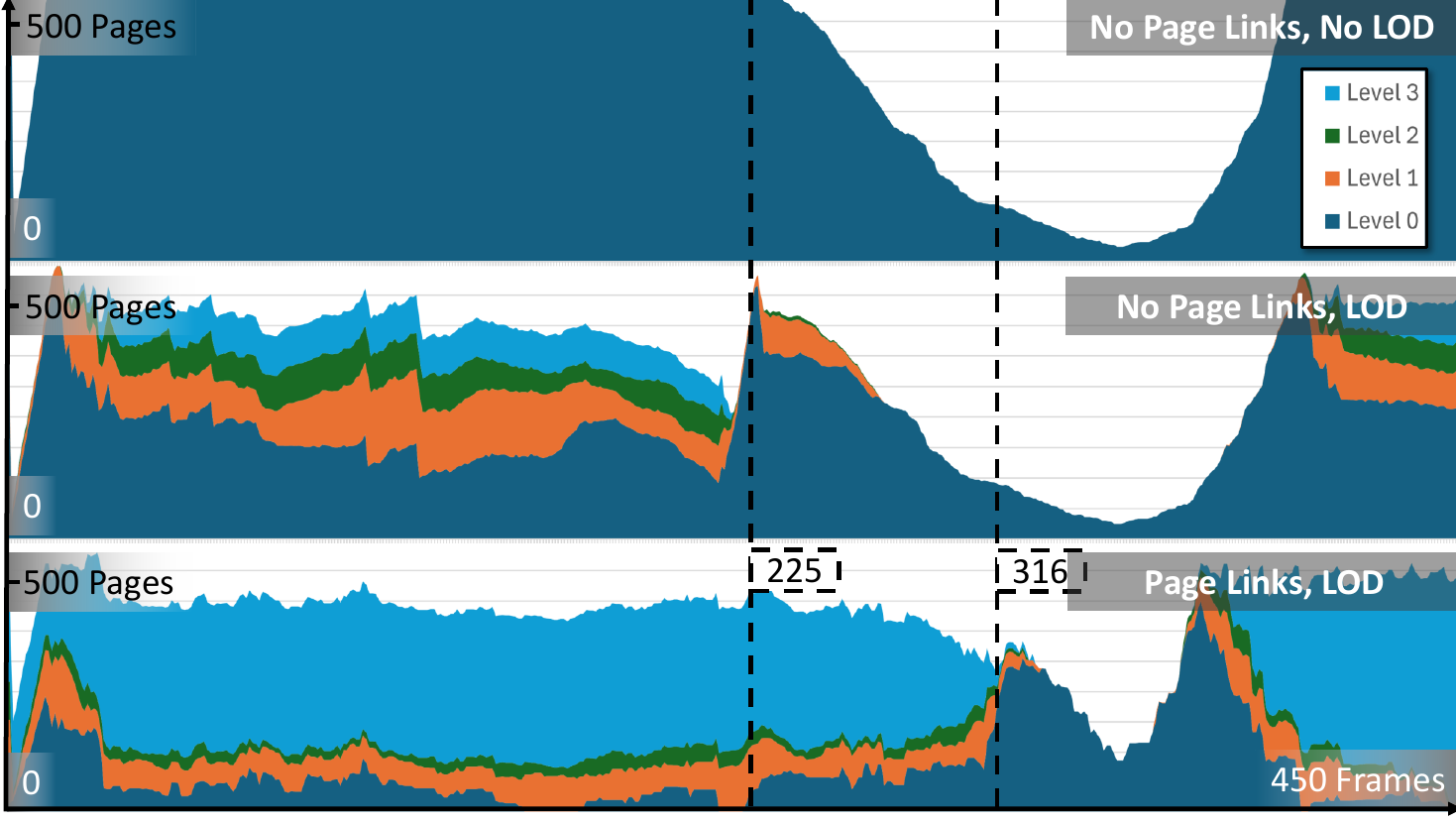}
    \caption{Memory usage during a fly-through in the scene "Residence" with different ablations. This scene contains almost 5000 pages of Gaussians with the first \gls{lod} level taking up more than 2 GiB. With a 500 page buffer only $\sim10\%$ of Gaussians can be rendered at a time.}
    \label{fig:memory_eval_1}
    \label{fig:memory_eval_1_a}
    \label{fig:memory_eval_1_b}
    \label{fig:memory_eval_1_c}
\end{figure}

\begin{figure}[t]
    \includegraphics[width=1\linewidth,trim=0 385 245 0, clip]{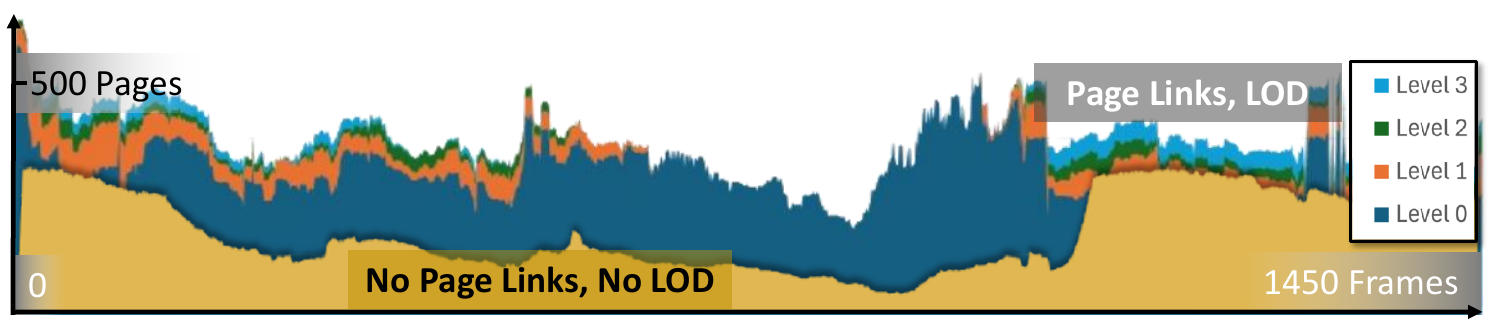}
    \caption{Memory usage during a fly-through in the scene "Alameda" with different ablations. This indoor scene contains many occlusions but only 730 pages in all.}
    \label{fig:memory_eval_2}
    \label{fig:memory_eval_2_a}
    \label{fig:memory_eval_2_b}
    \label{fig:memory_eval_2_c}
    %\postcapspace
\end{figure}

We assess memory usage for two scenes in more detail. The camera follows a path through each, with different ablations. As the camera moves, the number of rendered pages changes. In Residence (Figure~\ref{fig:memory_eval_1}), a large open scene, memory is quickly saturated without \gls{lod} enabled. In contrast, Alameda (Figure~\ref{fig:memory_eval_2}), which features many occlusions but is limited in size, memory use never exceeds 50\%. 

Upon enabling \gls{lod}, memory use in Residence is reduced drastically. A single physical page can store up to 2, 4, or 8 pages, respectively, at higher levels. Adaptive \gls{lod} attempts to keep memory use between 50\% and 80\% and modifies the distances at which levels transition to achieve this. %The effect is noticeable in Figure~\ref{fig:memory_eval_1_b}, with sharp spikes in memory use when it reaches either one of the limits. 
Alameda, which uses limited memory, is not affected by enabling \gls{lod}. All required pages fit within the 80\% limit in memory at all times, resulting in the use of level zero throughout.

When page links are used, memory usage is increased drastically. A large number of physical pages are populated with level three pages in Figure~\ref{fig:memory_eval_1_c}. This indicates adaptive \gls{lod} needs to set the distances for level transitions close to the camera in order to keep memory use within the targeted range. A similar increase in rendered pages can be observed in Figure~\ref{fig:memory_eval_2_c}, where multiple levels of detail are now required.

Page linking can indeed degrade image quality by increasing memory needs, leading to poorer rendering \gls{lod}. However, without page links scenes may show artifacts, as seen in Figure~\ref{fig:demo_lod_problems}.
%We can conclude that page linking can negatively affect image quality by increasing the required memory, which can cause pages to be rendered at a worse \gls{lod}. Unfortunately, the scene is susceptible to artifacts without page links, as observed in Figure~\ref{fig:demo_lod_problems}. %Figure~\ref{fig:demo_linking}
As a tradeoff, careful page assignments and setting a threshold to ignore small overlaps can minimize page links without causing significant artifacts.
%Greater care may be required when creating page assignments to reduce the number of page overlaps. Additionally, a threshold to ignore small overlaps may reduce the number of page links without introducing major artifacts.
Overall, using virtual memory significantly reduces memory usage and the count of Gaussians rendered. Without virtual memory, full scenes equate to 4417 and 724 pages. In the "Alameda" scene, visible Gaussians are at most 35\% of the total due to occlusions.
%However, in general using virtual memory can substantially decrease memory usage and the number of Gaussians rendered. The full scenes, if rendered without virtual memory, contain the equivalent of 4417 and 724 pages respectively. When rendering the scene "Alameda" with its many occlusions, visible Gaussians never exceed 35\% of all Gaussians in the scene.

\subsubsection{Performance}
\iffalse
\begin{figure}[t]
    \centering
    \includegraphics[width=\linewidth]{05/performance/performance_fps_pages_alameda_7}
    \precapspace
    \caption{\gls{fps} observed along a fixed camera path, compared to the number of rendered pages. \gls{fps} are not smoothed; they are determined by the time between subsequent frames. Performance clearly correlates with the number of Gaussians rendered. Frame times decrease as the number of pages decreases.}
    \postcapspace
    \label{fig:perf_fps_pages_alameda}
\end{figure}
\fi

\begin{figure}[t]
    \centering
    \includegraphics[width=\linewidth,trim=5 245 245 0,clip]{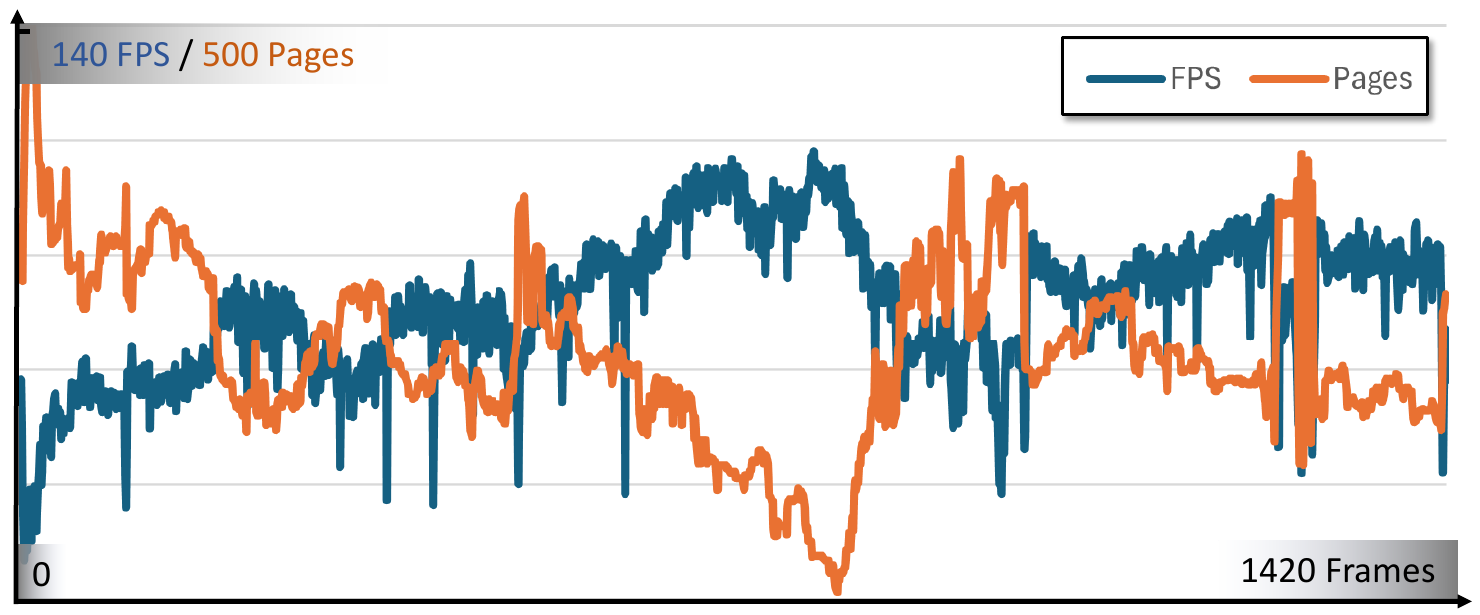}
    %\precapspace
    \caption{\gls{fps} observed along a fixed camera path, compared to the number of rendered pages. \gls{fps} are not smoothed; they are determined by the time between subsequent frames. Performance clearly correlates with the number of Gaussians rendered. Frame times decrease as the number of pages decreases.}
    %\postcapspace
    \label{fig:perf_fps_pages_alameda}
\end{figure}

\begin{table}[t]
	\centering
        \footnotesize
	\begin{tabular}{|p{2.4cm}||p{2.4cm}|p{2.4cm}|}
		\hline
		GPU & FP32 [TFLOPS] & Bandwidth [GB/s] \\
		\hline\hline
		NVIDIA GTX 1070 & 6.5~\cite{gtx1070} & 256.3~\cite{gtx1070} \\
		\hline
		Apple M1 \gls{soc} & 2.6~\cite{m1_flops} & 68.3~\cite{m1_memory} \\
		\hline
	\end{tabular}
	\caption{Basic comparison of the iPad Pro's integrated M1 \gls{gpu} to the dedicated \gls{gpu} used for evaluation on desktop. FP32 measures the number of 32-bit floating point operations per second (in trillions).}
	\label{tab:desktopvmobile}
    %\postcapspace
\end{table}

Performance, measured in \gls{fps}, is closely correlated with the number of Gaussians rendered (Figure~\ref{fig:perf_fps_pages_alameda}). As the number of pages increases, the \gls{fps} drop. This is indicative of the effectiveness of reducing the number of rendered Gaussians to improve performance. %The chart displays \gls{fps} based on the time between two subsequent frames. Values can be converted to the time taken in milliseconds using $\frac{1000}{FPS}$. It contains frame times between around 7 to 24 milliseconds.
%Figure~\ref{fig:performance_time} contains a more detailed breakdown of the time various steps take when rendering a frame. We list some frames of particular interest:
In the following we discuss selected frames from the Berlin scene with a certain impact: 
\begin{itemize}[leftmargin=*]
    \setlength\itemsep{0.0em}
    \footnotesize
    \item \textbf{Most Pages.} One of the frames with the highest number of physical pages in active use.
    \item \textbf{Median.} The median time taken to complete each step over the entire path. This is not a real frame but rather a reconstruction meant to approximate the median frame.
    \item \textbf{Shortest.} The frame with the smallest sum of the time taken by each step.
    \item \textbf{Largest Transfer.} The frame with the most bytes of Gaussian data copied to \gls{gpu} memory.
\end{itemize}

\iffalse
\begin{figure}[t]
	\centering
	\includegraphics[width=\linewidth]{05/performance/performance_zipnerf_berlin_4_times}
	\caption{Comparison of the time taken for various steps when rendering a frame with virtual memory. Specific frames are selected for further analysis. The median frame is not an actual frame but a combination of the median time taken by each step.}
	\label{fig:performance_time}
\end{figure}
\fi

\iffalse
\begin{figure}[t]
	\centering
	\includegraphics[width=\linewidth]{05/performance/performance_zipnerf_berlin_times_full}
	\caption{The time taken for various steps with virtual memory as well as the median and shortest frame rendered without virtual memory. The overhead associated with virtual memory is negligible when compared to the time gain for depth sort and render stages.}
	\label{fig:performance_time_full}
\end{figure}
\fi

\begin{figure}[t]
	\centering
	\includegraphics[width=\linewidth,trim=100 280 93 33,clip]{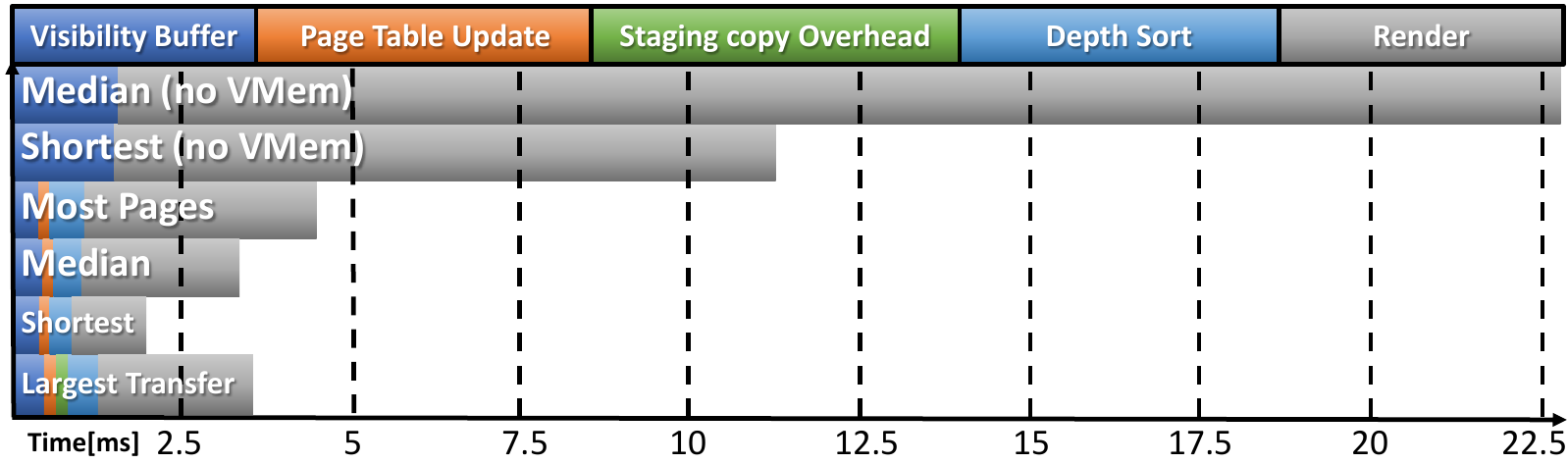}
	\caption{The time taken for various steps with virtual memory as well as the median and shortest frame rendered without virtual memory. The overhead associated with virtual memory is negligible when compared to the time gain for depth sort and render stages.}
	\label{fig:performance_time_full}
\end{figure}

\begin{figure}[t]
	\centering
	\includegraphics[width=\linewidth,trim=102 350 95 21,clip]{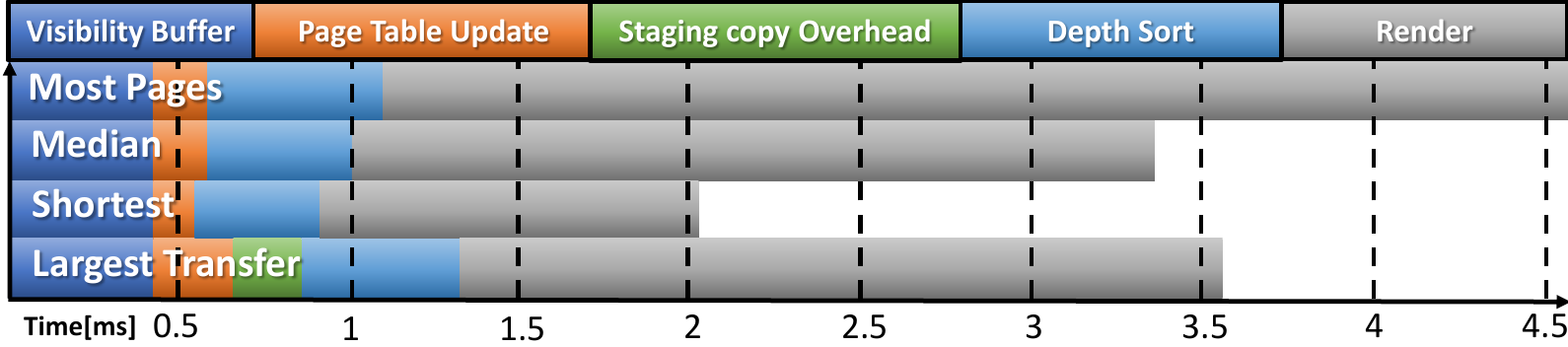}
	\caption{Comparison of the time taken for various steps when rendering a frame with virtual memory. Specific frames are selected for further analysis. The median frame is not an actual frame but a combination of the median time taken by each step.}
	\label{fig:performance_time}
    %\postcapspace
\end{figure}

Rendering the visibility buffer and reducing it to a list of page IDs takes an almost constant amount of time. %As described in Section~\ref{chap:implementation}, this time can be further reduced with additional optimizations. 
Page table updates are nearly constant, with a slight delay when adding many new pages. Most frames have no waiting overhead for the staging buffer copy. For the frame with the largest transfer, this is a minor issue. Sorting by depth and rendering time scale with the number of Gaussians, with rendering time being crucial to minimize.
%Page table updates are also near constant, only taking slightly longer when requiring many new pages. For most frames, there is no overhead from waiting for the copy to the staging buffer to be complete. In the case of the frame with the largest transfer, this is a minor factor. The time taken to both sort by depth and render the final frame clearly scales with the number of Gaussians. The time taken to render is especially crucial to minimize.

\iffalse
This is further illustrated by Figure~\ref{fig:performance_scatter}, where the effect the number of pages on depth sort and render times is explored. Clearly, as the number of pages increases, so does the time taken. Since all frames in the chart are rendered with virtual memory, the effect of memory locality is of little significance. 
\begin{figure}[t]
	\centering
	\includegraphics[width=\linewidth]{05/performance/performance_scatter_zipnerf_berlin_4}
	\caption{Relationship between the number of visible pages and the time taken to sort and render the contained Gaussians. As the number of required pages, and therefore the number of Gaussians that need to be sorted and rendered, increases so does the required time. This relationship holds although it may not be as apparent as it is in this chart with different settings and scales.}
	\label{fig:performance_scatter}
\end{figure}
\fi

The effectiveness of virtual memory can be seen when rendering the scene without virtual memory and comparing the timings (Figure~\ref{fig:performance_time_full}). The chart contains the timings shown in Figure~\ref{fig:performance_time} but adds additional data without the use of virtual memory. The rendered scene contains the same Gaussians as the base level of Gaussians with virtual memory. Gaussians are not grouped into pages, reordered, or padded. Therefore, it does not benefit from visibility determination or improved memory locality. 

Rendering the Berlin scene uses 489 pages in virtual memory, fitting comfortably in a 500-page buffer. Excluding hidden pages notably improves frame times.
%The scene in question (Berlin) would only take up 489 pages if it were rendered with virtual memory. Therefore, all pages could fit into the buffer with space for 500 pages at once. However, excluding hidden pages from rendering clearly makes a significant difference in frame times.
Frames without virtual memory take longer than previously analyzed ones, with the median frame taking even more time. Without virtual memory, page management overhead is eliminated. Timing shows the benefits of these preparatory steps, as even the depth sort stage is quicker. Rendering speeds up significantly due to improved memory locality and fewer Gaussians to process.
%Even the shortest frame without virtual memory dwarfs all of the previously analyzed frames. The median frame, which is again constructed from the median time taken for each step, takes even longer. Since neither of these use virtual memory, the overhead of page determination, page table management, and transfers are not a factor. Judging by the timings, the overhead introduced by these preparatory steps are clearly worth it in this scene. Notably, even the depth sort stage is shortened enough to make the overhead not a factor. Rendering is sped up dramatically, by a combination of better memory locality and a reduced number of Gaussians to render.

\subsubsection{Mobile}

Our concept renders discrete and integrated \gls{gpu}s identically under the same configuration and swapchain image size. However, based on Table~\ref{tab:desktopvmobile}, performance varies across devices.
%Our concept handles rendering for discrete and integrated \gls{gpu}s practically the same. Given the same configuration (limits, buffer sizes, etc.) and swapchain image size, the output is identical. However, given the numbers in Table~\ref{tab:desktopvmobile} we cannot expect the devices to perform the same.
The iPad's M1 is a \gls{soc} with \gls{cpu} and \gls{gpu} on a single chip, sharing the same memory. Vulkan reports just a single memory heap where memory may be accessed by both device and host. Our application makes no use of this, instead keeping the same data in memory multiple times. In an optimization file backed memory can be used by the \gls{gpu} directly and no manual streaming is necessary. After measuring timings we can therefore make some limited assumptions about performance without streaming.

With the same configuration and camera paths, only performance varies between desktop and mobile. Initially, comparing baseline performance without virtual memory, Figure~\ref{fig:performance_mobile_1} shows frames take longer on mobile due to a fourfold increase in Gaussians sorting time, indicating slower memory access as expected with an integrated \gls{gpu}. Enabling virtual memory reduces the Gaussians needing sorting. This method's overhead is justified, as depth sorting and rendering are significantly faster. However, mobile's virtual memory time is still much higher than desktop's, a gap a software renderer could likely reduce.
%With a matching configuration and camera paths only the performance differs between the desktop and mobile device. First, it is relevant to compare baseline performance of our renderer without virtual memory. Figure~\ref{fig:performance_mobile_1} indicates frames take slightly longer on mobile. This is mainly caused by a roughly four times increase in the time it takes to sort Gaussians by depth. This indicates slow memory accesses which is expected with an integrated \gls{gpu}. When virtual memory is enabled the number of Gaussians which need be sorted decrease. Overall, the overhead introduced by our method is clearly worth the time considering both depth sorting and rendering is much faster. The time taken with virtual memory on mobile still dwarfs the same on desktop. A software renderer can likely reduce this gap.

\iffalse
\begin{figure}[t]
	\centering
	\includegraphics[width=\linewidth]{05/performance/performance_mobile_1}
	\caption{Comparison of the time taken to render frames on desktop and mobile. All frames are a collection of the median times for each stage along a path. The first two bars are a baseline, without virtual memory enabled.}
	\label{fig:performance_mobile_1}
\end{figure}
\fi

\begin{figure}[t]
	\centering
	\includegraphics[width=\linewidth,trim=100 350 93 19,clip]{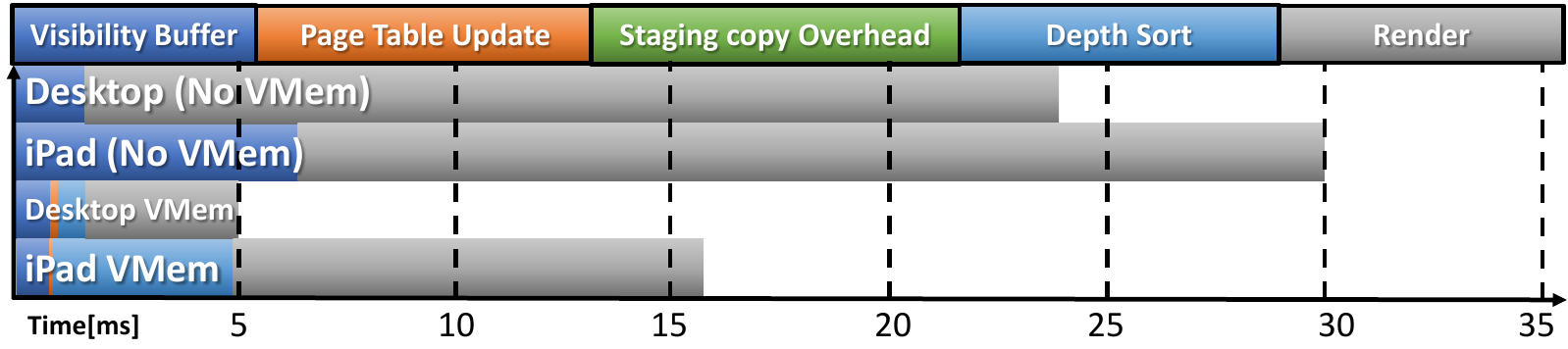}
	\caption{Comparison of the time taken to render frames on desktop and mobile. All frames are a collection of the median times for each stage along a path. The first two bars are a baseline, without virtual memory enabled.}
	\label{fig:performance_mobile_1}
\end{figure}

Figure~\ref{fig:performance_mobile_2} shows mobile-measured times for a larger scene with frame times over 200 ms without virtual memory. Our method reduces time by culling Gaussians, though streaming overhead remains due to slow memory access and lack of optimization for integrated \gls{gpu}s. We simulate median times without streaming overhead, potentially possible with file-backed memory. However, even then, it's not interactive. Large scenes need further optimization like reducing buffer sizes or software rendering.
%Figure~\ref{fig:performance_mobile_2} exclusively contains times measured on mobile. The scene is much larger which is clearly noticeable in the frame times of over 200ms without virtual memory. Our method culls enough Gaussians to reduce the time taken to about a quarter. This includes a sizable streaming overhead, again indicating slow memory accesses and the lack of optimization for integrated \gls{gpu}s. Finally, we construct a fictional frame of the median times without the overhead for streaming. This may be possible when using the file-backed memory for rendering directly. However, even in this best case scenario, fail to be interactive. For such large scenes additional optimizations are necessary. This may include reducing buffer sizes or implementing software rendering.

\iffalse
\begin{figure}[t]
	\centering
	\includegraphics[width=\linewidth]{05/performance/performance_mobile_2}
	\caption{Time taken to render a frame with the median of each stage on mobile. The last bar displays a fictional frame that is the same as the one above it but with the streaming overhead removed.}
	\label{fig:performance_mobile_2}
\end{figure}
\fi

\begin{figure}[t]
	\centering
	\includegraphics[width=\linewidth,trim=100 350 88 45,clip]{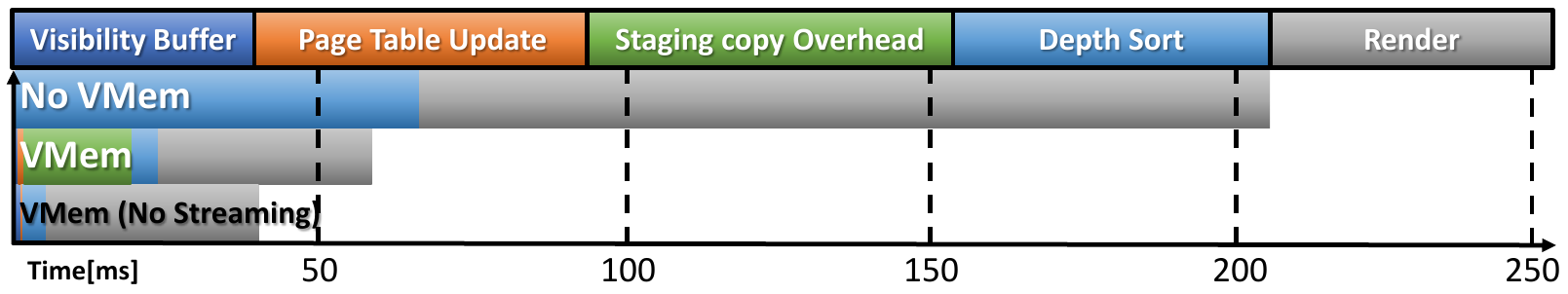}
	\caption{Time taken to render a frame with the median of each stage on mobile. The last bar displays a fictional frame that is the same as the one above it but with the streaming overhead removed.}
	\label{fig:performance_mobile_2}
    \postcapspace
\end{figure}

Overall, performance on mobile is mainly impacted by memory accesses. This is especially noticeable during our global depth sort. Our virtual memory method, apart from the unoptimized streaming, is not greatly impacted by the move to mobile.

\subsection{Discussion}
Our initial assumption is that the number of rendered Gaussians directly affects performance. For an approach to be effective, an induced reduction in render time must exceed the overhead of identifying visible Gaussians. This is clearly confirmed by our results. We expect this to also hold true with different rendering approaches, including the software renderer from Kerbl \etal~\cite{3dgs}. 

Our proxy mesh strategy effectively reduces memory in large scenes and relies on acceleration structures for fast Gaussian determination, as mentioned in Section~\ref{chap:related}. However, unlike other methods, it handles occlusions well, boosting performance in occluded environments such as indoor and city street-level datasets. Our approach also allows virtually unlimited scene size during rendering at a reasonable scale and speed.
%Visibility through a proxy mesh is a viable strategy. While memory requirements in large, open scenes can be reduced effectively with our method and recent methods (see Section~\ref{chap:related}), these often use acceleration structures for fast Gaussian determination in the viewport. However, they don't account for occlusions, making performance scene-dependent. Our method excels in heavily occluded scenes, like indoor environments or city street-level datasets, where accurately reconstructed walls and buildings block large scene areas. At reasonable scale and speed, scene size is virtually unlimited during rendering with our method.

Published datasets and implementations for large-scale scenes are inadequate. Common \gls{3dgs} training implementations poorly handle occlusions, creating artifacts. Datasets often originate from \gls{uav}s with few occlusions. The dataset from Kerbl \etal~\cite{hierarchical_3dgs} is promising but uses a non-standard file format.

Gaussians can't map directly to surface textures, so we use page links, which help avoid rendering artifacts but increase memory usage. We plan to enhance link creation to use fewer pages.
%Since Gaussians don't directly map to surfaces like textures, we use page links. This system involves a trade-off. Depending on the scene, they may be crucial for avoiding artifacts in rendered images. However, creating these links currently increases memory usage by requiring more pages. We aim to improve link creation methods to reduce their number in the future.
\gls{lod} integrates with our solution like mipmapping with virtual texturing, crucial for maintaining a strict memory budget without significant quality loss. Our \gls{lod} system has flaws in creation (incorrect scale) and rendering (lack of level blending). Integrating recent works from Section~\ref{chap:related} can address these issues.

% !TEX root = main.tex
\section{Conclusion and Future Work}
\label{chap:conclusions}

In this work, we demonstrate the viability of applying the concepts used in virtual texturing to cutting-edge research in \gls{3dgs}. 
A proxy mesh is generated from an existing \gls{3dgs} scene, grouping Gaussians into pages with IDs and linking them if overlapping, forming several \gls{lod} levels. Rendering this mesh to a visibility buffer enables fast visibility checks. Pages are managed in \gls{gpu} memory, transferring visible ones to the \gls{gpu} just in time, replacing unnecessary ones. The \gls{lod} level is selected based on camera distance and memory. Documented inspiration from virtual texturing, preprocessing uses a Python app with JIT optimizations, while a C++ app renders Gaussians in real-time with Vulkan API. We demonstrate its effectiveness in reducing Gaussians, minimizing memory and time for a \gls{3dgs} frame render.
% A proxy mesh is created from an existing \gls{3dgs} scene, where Gaussians are grouped into pages with marked IDs and linked if they overlap. Several \gls{lod} levels are formed by clustering Gaussians within a page. Rendering the proxy mesh to a visibility buffer facilitates fast visibility checks. A page table manages pages in \gls{gpu} memory, transferring visible pages to the \gls{gpu} just in time, replacing unnecessary ones. Depending on the camera distance and memory, an appropriate \gls{lod} level is selected. Our method, inspired by virtual texturing, is documented. Preprocessing uses a Python app with JIT optimizations, while a C++ app renders Gaussians in real-time via the Vulkan API. We evaluate and demonstrate the approach's effectiveness in reducing Gaussian numbers, minimizing memory and time for rendering a \gls{3dgs} frame.

We efficiently determine visible Gaussian pages by addressing occlusions with a mesh. While \gls{3dgs} involves millions of ellipsoids, making overlap-free grouping difficult, our method groups Gaussians by position. Page linking aligns mesh textures with Gaussians, preventing discrepancies. \gls{lod} shows reduced memory use and improved performance, opening research opportunities. We use virtual texturing to further cut memory and boost performance, vital for expanding \gls{3dgs} to larger scenes and aiding industry adoption, especially for mobile devices with integrated \gls{gpu}s and low hardware needs.
%We quickly determine which Gaussian pages are visible in a frame, surpassing existing methods by addressing occlusions within the view, using a mesh to implicitly detect them. \gls{3dgs} involves millions of ellipsoids, making neat, overlap-free grouping impossible, so our approach groups Gaussians by position. Page linking connects mesh surface textures with approximate Gaussians, avoiding discrepancies between proxy mesh and visible Gaussians. \gls{lod} demonstrates reduced memory needs and better performance, highlighting ongoing research opportunities.

In contrast to related works, we avoid compression in order to put additional stress on our system. Large scene reconstruction is still an active research topic and such tools are not readily available yet. Not compressing Gaussians allows us to test our method at its limits to effectively find shortcomings. We demonstrate the use of virtual texturing in this field, effectively reducing memory and enhancing performance. This is crucial for expanding \gls{3dgs} to larger scenes and promoting industry adoption, particularly for applications targeting mobile devices with integrated \gls{gpu}s and low hardware requirements.

In future work, several shortcomings are subject to further investigation. Our current page linking method corrects artifacts from Gaussians assigned to pages by their mean, but such overlaps are common, increasing the number of required pages. Future work might adjust page assignments to reduce overlaps, or ignore small overlaps without affecting image quality significantly. Our \gls{lod} solution, based on Yan \etal~\cite{anti_alias_lod} and mipmapping, serves as a proof of concept showing compatibility with virtual memory. However, it needs improvements to manage abrupt distance changes adaptively \gls{lod}, and the merging of Gaussians by a fixed factor doesn't cover them accurately. Popping artifacts during page level changes remain unaddressed. As seen in Section~\ref{chap:related}, recent research in \gls{lod} for \gls{3dgs} has gained interest; some approaches could complement our future work.

Our approach avoids Gaussian compression to explore its limitations. Enhanced datasets and algorithms will aid in testing large scenes. Future versions may adopt Gaussian compression through texture algorithms to reduce data copies and apply vector quantization for smaller codebooks in \gls{gpu}, lessening data transfer. Unlike Kerbl's \etal~\cite{3dgs} software rasterizer for \gls{3dgs} scenes, which sorts Gaussians by tile for potential performance gains, our method is hardware-based, sorting Gaussians by camera distance. Transparency challenges in page determination echo traditional virtual texturing issues (Mayer~\cite{vt_thesis}), crucial for realistic representations in \gls{3dgs} with transparencies. A possible solution is to pin pages on transparent surfaces in \gls{gpu}.
%Our concept avoids Gaussian compression to explore limitations. Improved datasets and reconstruction algorithms for large scenes will facilitate further testing. Future versions could integrate Gaussian compression via texture algorithms, reducing copied data, and use vector quantization for smaller codebooks in \gls{gpu} memory, minimizing data transfer. Kerbl's \etal~\cite{3dgs} original rendering for \gls{3dgs} scenes introduced a software rasterizer, unlike our hardware-based approach that sorts Gaussians by camera distance. The software rasterizer splits the image and sorts Gaussians per tile, a technique promising better performance, though performance trends from Section~\ref{chap:evaluation} are expected to remain with changes. Transparency challenges in page determination remain similar to traditional virtual texturing issues (Mayer~\cite{vt_thesis}), more crucial in \gls{3dgs} for realistic representation, which often involves transparencies. A potential fix involves pinning pages on transparent surfaces in \gls{gpu} memory.

Currently, our method avoids scene reconstruction changes, starting with a complete scene for preprocessing. Future work could integrate virtual memory into the training phase, regularly preprocessing and rendering with visibility determination.

%% if specified like this the section will be omitted in review mode
\acknowledgments{%
\footnotesize{Funded by the European Union under Grant Agreement No. 101092861. Views and opinions expressed are, however, those of the author(s) only and do not necessarily reflect those of the European Union or the European Commission. Neither the European Union nor the granting authority can be held responsible for them.}.%
}
\balance
\bibliographystyle{abbrv-doi-hyperref}

\bibliography{citations}

\appendix % You can use the `hideappendix` class option to skip everything after \appendix

\end{document}